\documentclass[]{elsarticle}

\usepackage{graphicx,color}
\usepackage{amssymb}
\usepackage{amsbsy}
\usepackage[normalem]{ulem}
\usepackage{amsmath,tikz}
\usetikzlibrary{calc,patterns, decorations.pathmorphing, decorations.markings}

\makeatletter
\newcommand{\vast}{\bBigg@{4}}
\newcommand{\Vast}{\bBigg@{5}}
\makeatother

\title{A principle of similarity for nonlinear vibration absorbers}

\author{G. Habib, G. Kerschen\\
email: giuseppe.habib@ulg.ac.be\\
Phone: +32 4 3664852\\
Space Structures and Systems Lab.\\
Aerospace and Mechanical Eng. Dept.\\
University of Li\`ege, Belgium
}

\begin{document}

\maketitle
Author pre-print. This paper is currently in preparation for publication in Physica D: Nonlinear Phenomena
\section*{Abstract}
\noindent{\it

This paper develops a principle of similarity for the design of a nonlinear absorber, the nonlinear tuned vibration absorber (NLTVA), attached to a nonlinear primary system. Specifically, for effective vibration mitigation, we show that the NLTVA should feature a nonlinearity possessing the same mathematical form as that of the primary system.
A compact analytical formula for the nonlinear coefficient of the absorber is then derived. 
The formula, valid for any polynomial nonlinearity in the primary system, is found to depend only on the mass ratio and on the nonlinear coefficient of the primary system. When the primary system comprises several polynomial nonlinearities, we demonstrate that the NLTVA obeys a principle of additivity, i.e., each nonlinear coefficient can be calculated independently of the other nonlinear coefficients using the proposed formula.}
\\

\noindent\textbf{Keywords}: principle of similarity, nonlinear resonances, equal-peak method, vibration absorber.

\section{Introduction}

\noindent

The use of linear resonators for the mitigation of resonant vibrations was first proposed by Watts \cite{Watts} and Frahm \cite{Frahm1909,Frahm1911} to reduce the rolling motion of ships. The problem was later formalized in more rigorous terms by Ormondroyd and Den Hartog \cite{Ormondroyd}, Den Hartog \cite{DenHartog} and Brock \cite{Brock}, who developed tuning rules that formed the basis of {\it Den Hartog's equal-peak method}. The vibration absorber considered in \cite{Ormondroyd,DenHartog,Brock} consists of a mass-spring-dashpot system attached to the primary system to be controlled. Through the proper tuning of the spring and damper of the absorber, it is possible to approximately obtain $H_{\infty}$ optimization of the frequency response in the vicinity of the target resonant frequency.

Thanks to its simplicity, effectiveness, low cost and small requirements for maintenance \cite{Fischer2007}, the passive vibration absorber (often referred to as tuned mass damper, tuned vibration absorber or dynamic vibration absorber) was extensively studied and implemented in real-life structures.
Its main applications include structures subject to human-induced vibrations, such as spectator stands and pedestrian bridges (the most famous example is the Millenium bridge in London \cite{Dallard}), steel structures excited by machines such as centrifuges and fans, aircraft engines \cite{Taylor1936}, helicopter rotors \cite{Hamouda1984}, tall and slender structures subject to wind-induced vibrations, but also power lines \cite{Sauter2002} and long-span suspended bridges \cite{Gu1998,Casalotti2014}. For a list of installations of vibration absorbers in civil structures the interested reader can refer to \cite{Holmes1995,Soto2013}.

An overview of existing designs for passive vibration absorbers is given in \cite{Fischer2007}. They include classical absorbers with translational mass movements, pendulum absorbers, centrifugal pendulum absorbers \cite{Mayet2014}, ball absorbers, sloshing liquid absorbers \cite{Housner} and particle vibration absorbers \cite{Lu2012}, although the sloshing liquid and particle vibration absorbers have qualitatively different features than the more typical vibration absorbers with a concentrated mass. Many different configurations and variations of the original vibration absorber were studied in the last decades, e.g., a damped host system \cite{Thompson}, different combinations of response and excitation \cite{Warburton} and the use of multi-vibration absorbers to control several resonances \cite{Kareem1995}.

A fundamental drawback of the vibration absorber (that hereafter is referred to as linear tuned vibration absorber, LTVA) is that it requires a fine tuning between its natural frequency and the targeted resonant frequency. The LTVA may therefore be ineffective when it is attached to nonlinear systems, in view of the frequency-amplitude dependence of these systems. This limitation was addressed in the literature through the development of nonlinear vibration absorbers \cite{Roberson,Nayfeh,Vakakis,Gendelman,Lacarbonara}. In \cite{HabibMSSP}, the authors introduced the nonlinear tuned vibration absorber (NLTVA), whose design is based on a principle of similarity with the nonlinear primary system, i.e., the NLTVA should be governed by equations similar to those of the primary system.
Thanks to this principle, a nonlinear generalization of Den Hartog's equal-peak method could be developed, and the optimal nonlinear coefficient of the NLTVA was determined numerically \cite{HabibMSSP}.
A more detailed study about the performance and robustness of the NLTVA was carried out in \cite{Detroux}, whereas it was experimentally validated in \cite{Grappasonni}. The NLTVA was also successfully applied for the mitigation of limit cycle oscillations \cite{HabibPRSA}.

This paper revisits the design procedure of the NLTVA proposed in \cite{HabibMSSP}. Specifically, by combining a harmonic balance technique with a perturbation method, it derives a compact analytical formula for the nonlinear coefficient of the absorber which is valid for any polynomial nonlinearity in the primary system. The case where the primary system comprises several nonlinearities is also carefully investigated in this study.

\section{Problem statement}

In this study, we seek to minimize the amplitude at resonance of a harmonically-forced one-degree-of-freedom (1DOF) nonlinear oscillator which models the targeted mode of the considered nonlinear system. This is achieved by attaching a nonlinear vibration absorber to the primary oscillator. Mathematically, the problem is formulated as follows:
\begin{equation}\label{ProbForm}
  \mbox{min} \|h(\omega)\|_\infty \rightarrow \mbox{min}\left\{\mbox{max}\left[|h(\omega_A)|,|h(\omega_B)|\right]\right\} \rightarrow |h\left(\omega_A\right)|=|h\left(\omega_B\right)|
\end{equation}
where $h(\omega)$ is the frequency response function of the coupled system measured at the primary mass, and $\omega_A,\omega_B$ represents the two resonance frequencies.

\subsection{The linear case}
\begin{figure*}
\begin{centering}
\includegraphics[trim = 10mm 10mm 5mm 10mm,width=0.45\textwidth]{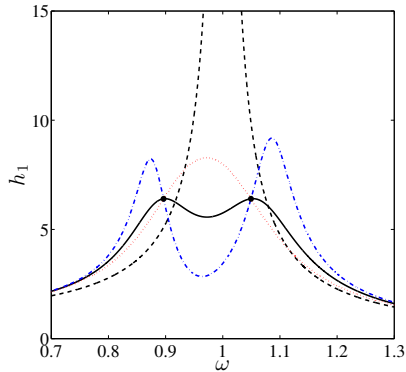}
\par\end{centering}
\caption{\label{linear}Frequency response function of an undamped linear primary system with an attached linear vibration absorber. Black dashed line: undamped primary system without absorber; blue dash-dotted and red dotted lines: absorber with damping smaller or greater than the optimal damping, respectively; black solid line: absorber with optimal stiffness and damping; black dots: invariant points.}
\end{figure*}

For linear absorbers coupled to linear oscillators, Den Hartog \cite{DenHartog} demonstrated that the frequency response function passes through two fixed points independent of absorber damping, and he selected the absorber stiffness that imposes equal amplitude for these points (Fig.~\ref{linear}).
Brock \cite{Brock} then calculated the absorber damping by taking the mean of the damping values that realize a maximum of the receptance at the two fixed points.
Even though the resulting formulas have sufficient accuracy in practice, they are only an approximation to Problem \ref{ProbForm}, because $|h\left(\omega_A\right)|$ is not strictly equal to $|h\left(\omega_B\right)|$.

Nishihara and Asami \cite{Nishihara,Asami} were the first to derive an exact solution to Problem \ref{ProbForm}. Instead of imposing two fixed points of equal amplitude, the direct minimization of the $h_{\infty}$ norm of the frequency response of the controlled structure was achieved. Eventually, exact analytical formulas were obtained for the frequency tuning $\lambda$ and damping ratios $\mu_2$:
\begin{eqnarray}\label{DHrule2}
\nonumber
    \lambda_{\text{opt}}&=&\frac{2}{1+\varepsilon}\sqrt{\frac{2\left(16+23\varepsilon+9\varepsilon^2+2(2+\varepsilon)\sqrt{4+3\varepsilon} \right)}
    {3(64+80\varepsilon+27\varepsilon^2)}}\\
    \mu_{2,\text{opt}}&=&\frac{1}{4}\sqrt{\frac{8+9\varepsilon-4\sqrt{4+3\varepsilon}}{1+\varepsilon}}
\end{eqnarray}
where $\varepsilon=m_2/m_1$ is the mass ratio ($\lambda$ and $\mu_2$ are defined later). The resonance frequencies are
\begin{equation}
\begin{split}\label{ResFreq}
\omega_A,\omega_B&=\Bigg[\frac{1}{2}\left(1+ (1 + \varepsilon) \lambda_{\text{opt}}^2 - 2(1 + \varepsilon)^2 \lambda_{\text{opt}}^2 \mu_{2,\text{opt}}^2\right)\\
&\mp \sqrt{\frac{1}{4}\left(1 + (1 + \varepsilon) \lambda_{\text{opt}}^2 - 2(1 + \varepsilon)^2 \lambda_{\text{opt}}^2 \mu_{2,\text{opt}}^2\right)^2 - \lambda_{\text{opt}}^2 r}\Bigg]^{1/2}
\end{split}
\end{equation}
where $r=8 \left((4 + 3 \varepsilon)^{3/2} - \varepsilon\right)/\left(64 + 80 \varepsilon + 27 \varepsilon^2\right)$.

We note that expressions (\ref{DHrule2}) and (\ref{ResFreq}) are valid for an undamped primary system.

\subsection{The nonlinear case}

\begin{figure}
\begin{centering}
\begin{tikzpicture}[scale=1.2]
\tikzstyle{spring}=[thick,decorate,decoration={zigzag,pre length=0.3cm,post length=0.3cm,segment length=6}]
\tikzstyle{damper}=[thick,decoration={markings,
  mark connection node=dmp,
  mark=at position 0.5 with
  {
    \node (dmp) [thick,inner sep=0pt,transform shape,rotate=-90,minimum width=15pt,minimum height=3pt,draw=none] {};
    \draw [thick] ($(dmp.north east)+(2pt,0)$) -- (dmp.south east) -- (dmp.south west) -- ($(dmp.north west)+(2pt,0)$);
    \draw [thick] ($(dmp.north)+(0,-5pt)$) -- ($(dmp.north)+(0,5pt)$);
  }
}, decorate]
%
\draw[thick] (0,1)--(0,-1);
\draw[spring] (0,.5)--(2,.5);
\node[above left] at (1,.5) {$k_{11}$};
\draw[spring] (0,-.5)--(2,-.5);
\draw[thick,->] (.5,-0.9)--(1.7,-.2);
\node[above left] at (1,-.5) {$\sum k_{1i}$};
\draw (2,1)rectangle(4,-1);
\node at (3,0) {$m_1$} ;
\draw[->] (3,1)|-(3.5,1.5);
\node[above] at (3.5,1.5) {$x_1$};
\draw[thick,->] (1.8,1.2)--(2.5,1.2);
\node[above] at (2,1.2) {$f\cos\omega t$};
\draw[spring] (4,.7)--(6,.7);
\node[above left] at (5,.7) {$k_{21}$};
\draw[damper] (4,0)--(6,0);
\node[above left] at (5,0) {$c_2$};
\draw[spring] (4,-.7)--(6,-.7);
\node[above left] at (5,-.7) {$\sum k_{2i}$};
\draw[thick,->] (4.5,-1.1)--(5.7,-.4);
\draw (6,0.8)rectangle(7,-.8);
\node at (6.5,0) {$m_2$} ;
\draw[->] (6.5,.8)|-(7,1.3);
\node[above] at (7,1.3) {$x_2$};
\end{tikzpicture}
\par\end{centering}
\caption{\label{model}Mechanical model.}
\end{figure}
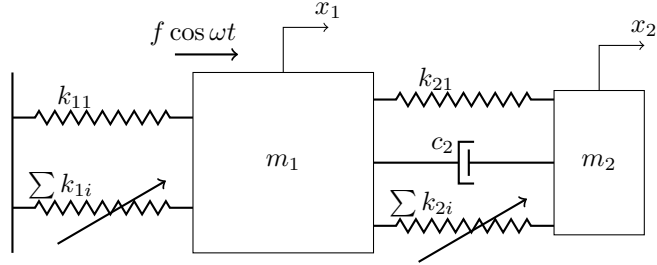

We now consider a nonlinear primary system with a polynomial restoring force symmetric with respect to the origin to which a nonlinear vibration absorber is attached, as illustrated in Fig.~\ref{model}.
Following the idea of the NLTVA developed in \cite{HabibMSSP,HabibPRSA}, the absorber should possess a restoring force characterized by the same mathematical form as that of the primary system, thus obeying a `principle of similarity'. The equations of motion of the coupled system are
\begin{equation}
\begin{split}
&m_1\ddot x_1+k_{11}x_1+\sum_{i=2}^nk_{1i}\left|x_1^i\right|\text{sign}\left(x_1\right)+c_2\left(\dot x_1-\dot x_2\right)\\
&\qquad+k_{21}\left(x_1-x_2\right)+\sum_{i=2}^nk_{2i}\left|x_1-x_2\right|^i\text{sign}\left(x_1-x_2\right)=f\cos\omega t\\
&m_2\ddot x_2+c_2\left(\dot x_2-\dot x_1\right)+k_{21}\left(x_2-x_1\right)+\sum_{i=2}^nk_{2i}\left|x_2-x_1\right|^i\text{sign}\left(x_2-x_1\right)=0
\end{split}\label{model0}
\end{equation}
where $x_1$ and $x_2$ are the displacements of the primary system and of the NLTVA, respectively; $m_1$, $m_2$, $k_{11}$ and $k_{21}$ are the masses and the linear spring coefficients of the primary system and of the NLTVA, respectively; $c_2$ is the damping coefficient of the NLTVA; $k_{1i}$ and $k_{2i}$, $i=2,...,n$, are the nonlinear spring coefficients of the primary system and of the NLTVA, respectively; $f$ and $\omega$ are the forcing amplitude and frequency, respectively, and $n$ is the highest order of nonlinearity present in the primary system.

Defining the dimensionless time $\tau=t\omega_{n1}=t\sqrt{k_{11}/m_1}$ and introducing the variables $q_1=x_1k_{11}/f$ and $q_2=x_2k_{11}/f$, the system is transformed into
\begin{equation}
\begin{split}
&q_1''+q_1+2\mu_2\lambda\varepsilon\left(q_1'-q_2'\right)+\lambda^2\varepsilon\left(q_1-q_2\right)+\sum_{i=2}^n\alpha_i\left|q_1^i\right|\text{sign}\left(q_i\right) \\
&\qquad+\varepsilon\sum_{i=2}^n b_i\alpha_i\left|q_1-q_2\right|^i\text{sign}\left(q_1-q_2\right)=\cos\gamma\tau\\
&\varepsilon q_2''+2\mu_2\lambda\varepsilon\left(q_2'-q_1'\right)+\lambda^2\varepsilon\left(q_2-q_1\right)+\varepsilon\sum_{i=3,5,\hdots}^nb_i\alpha_i\left(q_2-q_1\right)^i \\
&\qquad+\varepsilon\sum_{i=2}^nb_i\alpha_i\left|q_2-q_1\right|^i\text{sign}\left(q_2-q_1\right)=0\label{dimless}
\end{split}
\end{equation}
where $\mu_2=c_2/\left(2\sqrt{m_2k_{21}}\right)$, $\lambda=\omega_{n2}/\omega_{n1}=\sqrt{k_{21}m_1/(k_{11}m_2)}$, $\gamma=\omega/\omega_{n1}$, $\alpha_i=k_{1i}f^{i-1}/k_{11}^i$ and $b_i=k_{2i}/\left(\varepsilon k_{1i} \right)$.
The prime indicates derivation with respect to the dimensionless time $\tau$.

In the dimensionless system (\ref{dimless}), the forcing amplitude appears only in the expression of the nonlinear coefficients, which means that it is equivalent to consider a strongly nonlinear system or a system with a large forcing amplitude. In addition, the forcing amplitude modifies linearly the quadratic coefficients, quadratically the cubic coefficients, etc. This suggests that, if an optimal set of absorber parameters is chosen for a specific value of $f$, variations of $f$ will detune the NLTVA, unless the nonlinear coefficients of the primary system and of the absorber undergo a similar variation with $f$. This observation justifies the so-called principle of similarity.

Considering that the mass ratio $\varepsilon$ is imposed by practical constraints, the objective of this paper is to obtain analytical expressions of the NLTVA parameters, i.e., $\lambda$, $\mu_2$ and $\mathbf b\equiv [b_2,...,b_n]$, that realize equal peaks in the nonlinear frequency response of the primary oscillator for an as large as possible range of forcing amplitudes. This study therefore generalizes the formula obtained for a cubic primary oscillator in \cite{HabibMSSP}.

\section{Analytical extension of the equal-peak method to nonlinear systems}\label{procedure}

To ensure equal peaks at low energy levels for which the nonlinearities in the primary system are not activated, the linear parameters $\lambda$ and $\mu_2$ of the NLTVA are calculated as in the linear case, i.e., using Equations (\ref{DHrule2}).

The nonlinear coefficient vector $\mathbf{b}$ is determined in this section by extending the procedure proposed by Asami and Nishihara \cite{Asami} to the nonlinear case. To this end, the analytical procedure combines a harmonic balance technique and a perturbation method as in \cite{Luongo2012}, but the multiple scales method is replaced herein by a series expansion that considers small values of parameters $\alpha_i$. Eventually, we will show that this assumption does not limit the validity of the developments to weakly nonlinear regimes of motion thanks to the adoption of the principle of similarity.

\subsection{Approximate solution of the nonlinear problem}

In order to transform the system of nonlinear differential equations into a system of nonlinear algebraic equations, $q_1$ and $q_2$ are expanded in Fourier series \begin{equation}
\begin{split}
q_1\approx A_0+\sum_{i=1}^mA_{1i}\cos(i\gamma\tau)+\sum_{i=1}^mA_{2i}\sin(i\gamma\tau)&\\
q_2\approx B_0+\sum_{i=1}^mB_{1i}\cos(i\gamma\tau)+\sum_{i=1}^mB_{2i}\sin(i\gamma\tau)&.
\end{split}
\end{equation}
Regardless of the chosen maximal harmonic $m$, the system can be expressed in the form \begin{equation}
\mathbf{W}(\gamma)\mathbf y+\sum_{i=2}^n\alpha_i\left(\mathbf d_{i0}(\mathbf y)+b_i\mathbf d_{i1}(\mathbf y)\right)=\mathbf c,\label{dopoHB}
\end{equation}
where $\mathbf W$ is related to the linear part of the system, $\mathbf y$ collects the amplitude of the different harmonics of the solution, $\mathbf d_{i0}$ and $\mathbf d_{i1}$ contain the nonlinear terms and $\mathbf c$ is related to external forcing. For example, in the case of a cubic nonlinearity in the primary system and limiting the analysis to a single harmonic, i.e., $\cos^3\left(\gamma\tau\right)\approx 3/4\cos\left(\gamma\tau\right)$ and $\sin^3\left(\gamma\tau\right)\approx 3/4\sin\left(\gamma\tau\right)$, we obtain
\begin{small}
\begin{equation}
\begin{split}
&\left[\begin{array}{cccc}
1+\lambda^2\varepsilon-\gamma^2&2\mu_2\lambda\varepsilon\gamma &-\lambda^2\varepsilon &-2\mu_2\lambda\varepsilon\gamma\\
-2\mu_2\lambda\varepsilon\gamma & 1+\lambda^2\varepsilon-\gamma^2 & 2\mu_2\lambda\varepsilon\gamma & -\lambda^2\varepsilon\\
-\lambda^2\varepsilon&-2\mu_2\lambda\varepsilon\gamma&\lambda^2\varepsilon-\gamma^2\varepsilon& 2\mu_2\lambda\varepsilon\gamma\\
2\mu_2\lambda\varepsilon\gamma & -\lambda^2\varepsilon & -2\mu_2\lambda\varepsilon\gamma & \lambda^2\varepsilon-\gamma^2\varepsilon
\end{array}
\right]
\left[\begin{array}{c}A_{11}\\A_{21}\\B_{11}\\B_{21}\end{array}\right]\\
&\quad+\frac{3}{4}\alpha_3\Vast(\left[\begin{array}{c}
A_{11} \left(A_{11}^2+A_{21}^2\right)\\
A_{21} \left(A_{11}^2+A_{21}^2\right)\\
0\\
0\end{array}\right]\\
&\quad+\frac{3}{4}b_3\varepsilon\left[\begin{array}{c}
(A_{11}-B_{11}) \left(A_{11}^2-2 A_{11} B_{11}+A_{21}^2-2 A_{21} B_{21}+B_{11}^2+B_{21}^2\right)\\
(A_{21}-B_{21}) \left(A_{11}^2-2 A_{11} B_{11}+A_{21}^2-2 A_{21} B_{21}+B_{11}^2+B_{21}^2\right)\\
(B_{11}-A_{11}) \left(A_{11}^2-2 A_{11} B_{11}+A_{21}^2-2 A_{21} B_{21}+B_{11}^2+B_{21}^2\right)\\
(B_{21}-A_{21}) \left(A_{11}^2-2 A_{11} B_{11}+A_{21}^2-2 A_{21} B_{21}+B_{11}^2+B_{21}^2\right)\end{array}\right]\Vast)=\left[\begin{array}{c}1\\0\\0\\0\end{array}\right]. \label{example}
\end{split}
\end{equation}\end{small}

Considering that $\alpha_i$ are small parameters, we expand $\mathbf y$ with respect to $\alpha_i$ at the first order \begin{equation}
\mathbf y\approx\mathbf y_0+\sum_{i=2}^n\alpha_i\mathbf y_i,
\end{equation}
thus obtaining from Eq.~(\ref{dopoHB})
\begin{equation}
\mathbf{Wy}_0+\sum_{i=2}^n\alpha_i\left(\mathbf{Wy}_i+\left(\mathbf d_{i0}+b_i\mathbf d_{i1}\right)|_{\mathbf y=\mathbf y_0}\right)=\mathbf c.\label{dopoDec}
\end{equation}
Decomposing Eq.~(\ref{dopoDec}) with respect to the different parameters $\alpha_i$, the vectors $\mathbf y_i$ can be explicitly calculated through the formulas
\begin{equation}
\begin{split}
\mathbf{Wy}_0=\mathbf c&\rightarrow \mathbf y_0=\mathbf W^{-1}\mathbf c\\
\mathbf{Wy}_i+\left(\mathbf d_{i0}+b_i\mathbf d_{i1}\right)|_{\mathbf y=\mathbf y_0}=\mathbf 0&\rightarrow \mathbf y_i=-\mathbf W^{-1}\left(\mathbf d_{i0}+b_i\mathbf d_{i1}\right)|_{\mathbf y=\mathbf y_0}.
\end{split}
\end{equation}
The frequency response function $h(\gamma)$, which describes the maximal value of $q_1$ for different forcing frequencies and which is the key quantity in the equal-peak method, can be identified from the amplitude of the different harmonics contained in $\mathbf y$. For practical convenience, we consider the square of the frequency response $H=h^2$ and substitute the parameter $\gamma$ with its square $\Gamma=\gamma^2$.
Neglecting the higher-order terms of $\alpha_i$, the square of the frequency response takes the form
\begin{equation}
H=H_0+\sum_{i=2}^n\alpha_i\left(H_{i0}+b_iH_{i1}\right),\label{approx}
\end{equation}
where $H_0$, $H_{i0}$ and $H_{i1}$ are obtained analytically, but they are complicated functions of $\Gamma$.

\subsection{Identification of the nonlinear resonant frequencies}

\noindent Considering the nonlinear system, the square of the resonant frequencies $\Gamma_A$ and $\Gamma_B$ is given by the zeros of $\partial_{\,\Gamma}H=0$, where the $\partial_{\,\Gamma}$ indicates derivation with respect to $\Gamma$.
Since we consider small values of $\alpha_i$, $i=2,...,n$, $\Gamma_A$ and $\Gamma_B$ can be considered as small variations of their linear counterparts in Eq.~(\ref{ResFreq}), $\hat\Gamma_A=\omega_A^2$ and $\hat\Gamma_B=\omega_B^2$.
$\Gamma_A$ and $\Gamma_B$ are thus obtained linearizing $\partial_{\,\Gamma} H$ around $\hat\Gamma_A$ and $\hat\Gamma_B$, respectively.

For $\Gamma_A$ we have \begin{equation}
\partial_{\,\Gamma} H\approx \partial_{\,\Gamma} H|_{\Gamma=\hat\Gamma_A}+\partial^2_\Gamma H|_{\Gamma=\hat\Gamma_A}\left(\Gamma-\hat\Gamma_A\right),
\end{equation}
where $\partial_{\,\Gamma} H|_{\Gamma=\hat\Gamma_A}$ and $\partial^2_\Gamma H|_{\Gamma=\hat\Gamma_A}$ can be explicitly calculated through the relations \begin{equation}
\begin{split}
\partial_{\,\Gamma} H|_{\Gamma=\hat\Gamma_A}&=\partial_{\,\Gamma} H_0|_{\Gamma=\hat\Gamma_A}+\sum_{i=2}^n\alpha_i\left(\partial_{\,\Gamma} H_{i0}+b_i\partial_{\,\Gamma} H_{i1}\right)|_{\Gamma=\hat\Gamma_A}\\
\partial^2_\Gamma H|_{\Gamma=\hat\Gamma_A}&=\partial^2_\Gamma H_0|_{\Gamma=\hat\Gamma_A}+\sum_{i=2}^n\alpha_i\left(\partial^2_\Gamma H_{i0}+b_i\partial^2_\Gamma H_{i1}\right)|_{\Gamma=\hat\Gamma_A},
\end{split}
\end{equation}
where $\partial_{\,\Gamma} H_0|_{\Gamma=\hat\Gamma_A}=0$.
Imposing $\partial_{\,\Gamma} H=0$ we have \begin{equation}
\delta_A=\Gamma_A-\hat\Gamma_A\approx -\frac{\partial_{\,\Gamma} H}{\partial^2_\Gamma H}\bigg|_{\Gamma=\hat\Gamma_A}.
\end{equation}
Linearizing $\delta_A$ with respect to $\alpha_i=0$, $i=2,...,n$, we have \begin{equation}
\delta_A\approx -\sum_{i=2}^n\frac{\partial_{\,\Gamma} H_{i0}+b_i\partial_{\,\Gamma} H_{i1}}{\partial^2_\Gamma H_0}\bigg|_{\Gamma=\hat\Gamma_A}\alpha_i\label{deltaA}
\end{equation}
and analogously for $\Gamma_B$ \begin{equation}
\delta_B=\Gamma_B-\hat\Gamma_B\approx -\sum_{i=2}^n\frac{\partial_{\,\Gamma} H_{i0}+b_i\partial_{\,\Gamma} H_{i1}}{\partial^2_\Gamma H_{0}}\bigg|_{\Gamma=\hat\Gamma_B}\alpha_i.\label{deltaB}
\end{equation}

As it will be shown in Eq.~(\ref{bi}), the knowledge of $\delta_A$ and $\delta_B$ is not required for calculating $\mathbf b$.

\subsection{Definition of the optimal nonlinear coefficients $b_i$}

\noindent The equal-peak condition is verified if and only if the objective function \begin{equation}
F=H|_{\Gamma=\Gamma_A}-H|_{\Gamma=\Gamma_B}=0.\label{obj_funt}
\end{equation}
This condition is satisfied for the underlying linear system if $\lambda$ and $\mu_2$ are chosen according to Eq.~(\ref{DHrule2}).

Expanding $H$ in Taylor series around $\hat\Gamma_A$ and $\hat\Gamma_B$, $F$ becomes \begin{equation}
\begin{split}
F=&H|_{\Gamma=\hat\Gamma_A}+\partial_{\,\Gamma} H|_{\Gamma=\hat\Gamma_A}\delta_A+O(\delta_A^2)-H|_{\Gamma=\hat\Gamma_B}-\partial_{\,\Gamma} H|_{\Gamma=\hat\Gamma_B}\delta_B+O(\delta_B^2)\approx\\
\approx&\sum_{i=2}^n\alpha_i\left(H_{i0}|_{\Gamma=\hat\Gamma_A}+b_iH_{i1}|_{\Gamma=\hat\Gamma_A}\right)\\
&+\left(\partial_{\,\Gamma} H_0|_{\Gamma=\hat\Gamma_A}+\sum_{i=2}^n\alpha_i\left(\partial_{\,\Gamma} H_{i0}|_{\Gamma=\hat\Gamma_A}+b_i\partial_{\,\Gamma} H_{i1}|_{\Gamma=\hat\Gamma_A}\right)\right)\delta_A\\
&-\sum_{i=2}^n\alpha_i\left(H_{i0}|_{\Gamma=\hat\Gamma_B}+b_iH_{i1}|_{\Gamma=\hat\Gamma_B}\right)-\\
&\left(\partial_{\,\Gamma} H_0|_{\Gamma=\hat\Gamma_B}+\sum_{i=2}^n\alpha_i\left(\partial_{\,\Gamma} H_{i0}|_{\Gamma=\hat\Gamma_B}+b_i\partial_{\,\Gamma} H_{i1}|_{\Gamma=\hat\Gamma_B}\right)\right)\delta_B=0.
\end{split}\label{F}
\end{equation}
Since $\delta_A$ and $\delta_B$ are, in first approximation, of the order $O\left(\alpha_i\right)$ and since $\partial_{\,\Gamma} H_0|_{\Gamma=\hat\Gamma_A}=\partial_{\,\Gamma} H_0|_{\Gamma=\hat\Gamma_B}=0$, and limiting the analysis to terms of first order, Eq.~(\ref{F}) reduces to 
\begin{equation}
F\approx\sum_{i=2}^n\alpha_i\left(H_{i0}|_{\Gamma=\hat\Gamma_A}+b_iH_{i1}|_{\Gamma=\hat\Gamma_A}-H_{i0}|_{\Gamma=\hat\Gamma_B}-b_iH_{i1}|_{\Gamma=\hat\Gamma_B}\right)=0.\label{stepbi}
\end{equation}
Decomposing Eq.~(\ref{stepbi}) with respect to $\alpha_i$, $i=2,...,n$, and solving with respect to $\mathbf b$, we have
\begin{equation}
b_i=\frac{\left.H_{i0}\right|_{\Gamma=\hat\Gamma_A}-\left.H_{i0}\right|_{\Gamma=\hat\Gamma_B}}
{-\left.H_{i1}\right|_{\Gamma=\hat\Gamma_A}+\left.H_{i1}\right|_{\Gamma=\hat\Gamma_B}}\label{bi}.
\end{equation}

The coefficients $b_i$, as expressed in Eq.~(\ref{bi}), depend only on the order of nonlinearity under consideration $i$ and on the linear terms.
Since the coefficients of the linear terms are fully identified by $\varepsilon$, $b_i$ are function of $\varepsilon$ only.
They can be calculated from the knowledge of the approximated frequency response, as defined in Eq.~(\ref{approx}), for $\Gamma=\hat\Gamma_A$ and $\Gamma=\hat\Gamma_B$, without requiring any further information about the system.
If $H(\Gamma)$ is kept in its analytical form, the coefficients $\mathbf b$ can be defined analytically through computer algebra using Eq.~(\ref{bi}). However, even considering a system with a single cubic nonlinearity and limiting the analysis to a single harmonic, the final formula expressing $b_3$ is extremely long.

An important theoretical result of the outlined procedure is that, in first approximation, there is no interaction between the polynomial nonlinearities. This means that, if different polynomial nonlinearities are present in the primary system, the polynomial nonlinearities in the NLTVA can be designed independently of each other. Eventually, they can be simply summed up according to an {\it additivity property}. 
This finding greatly simplifies the design of the NLTVA and is verified numerically in Section \ref{addit}. We note that additivity of nonlinear components was also observed for a multi-degree-of-freedom nonlinear energy sink in \cite{Vaurigaud}.

\section{Proposed tuning rule for the NLTVA}\label{rule}

\subsection{Complete analytical design of the NLTVA}
\begin{figure*}
\begin{centering}
\makebox[\textwidth][c]{\includegraphics[trim = 10mm 10mm 5mm 10mm,width=0.39\textwidth]{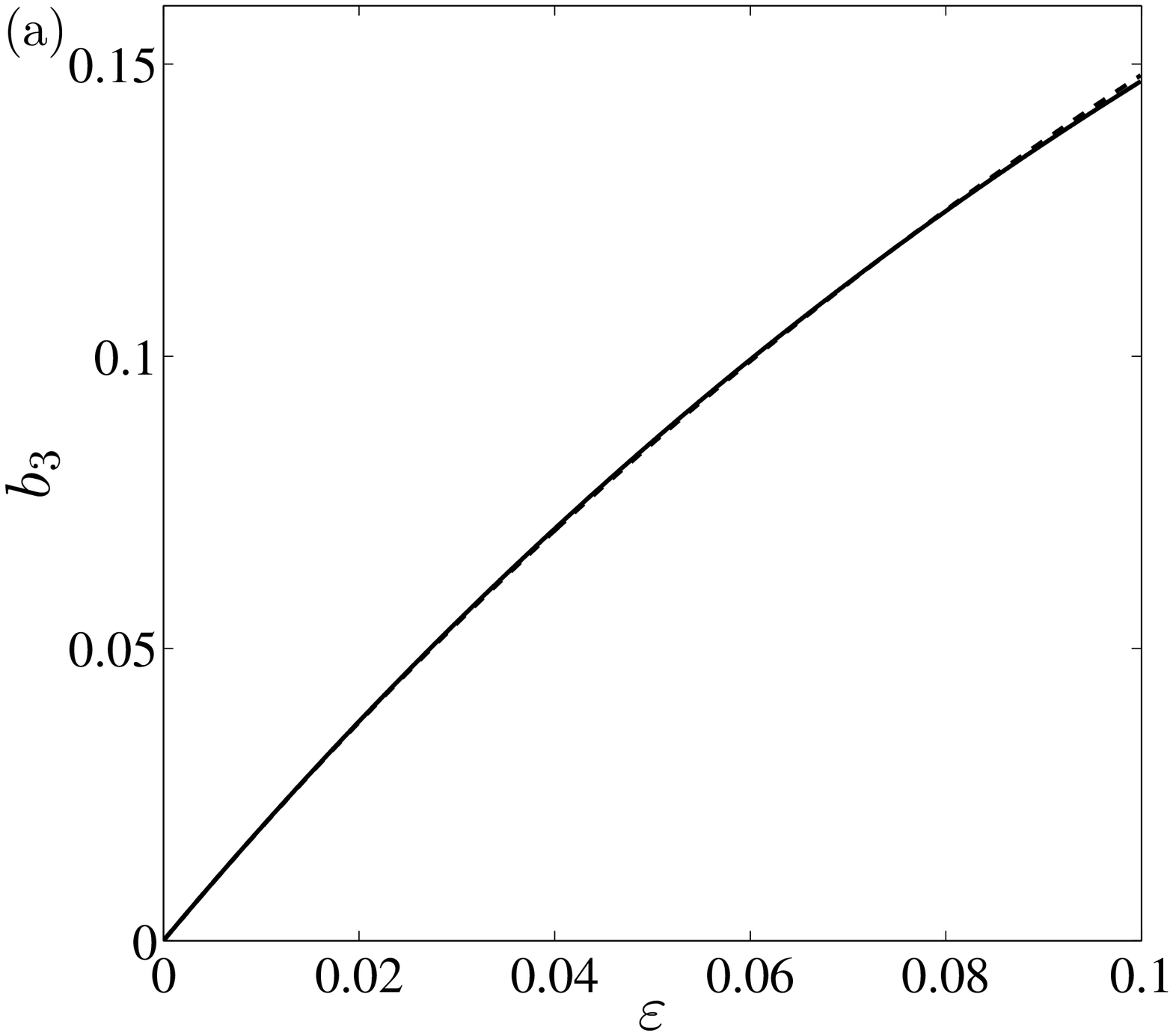}
\includegraphics[trim = 10mm 10mm 5mm 10mm,width=0.39\textwidth]{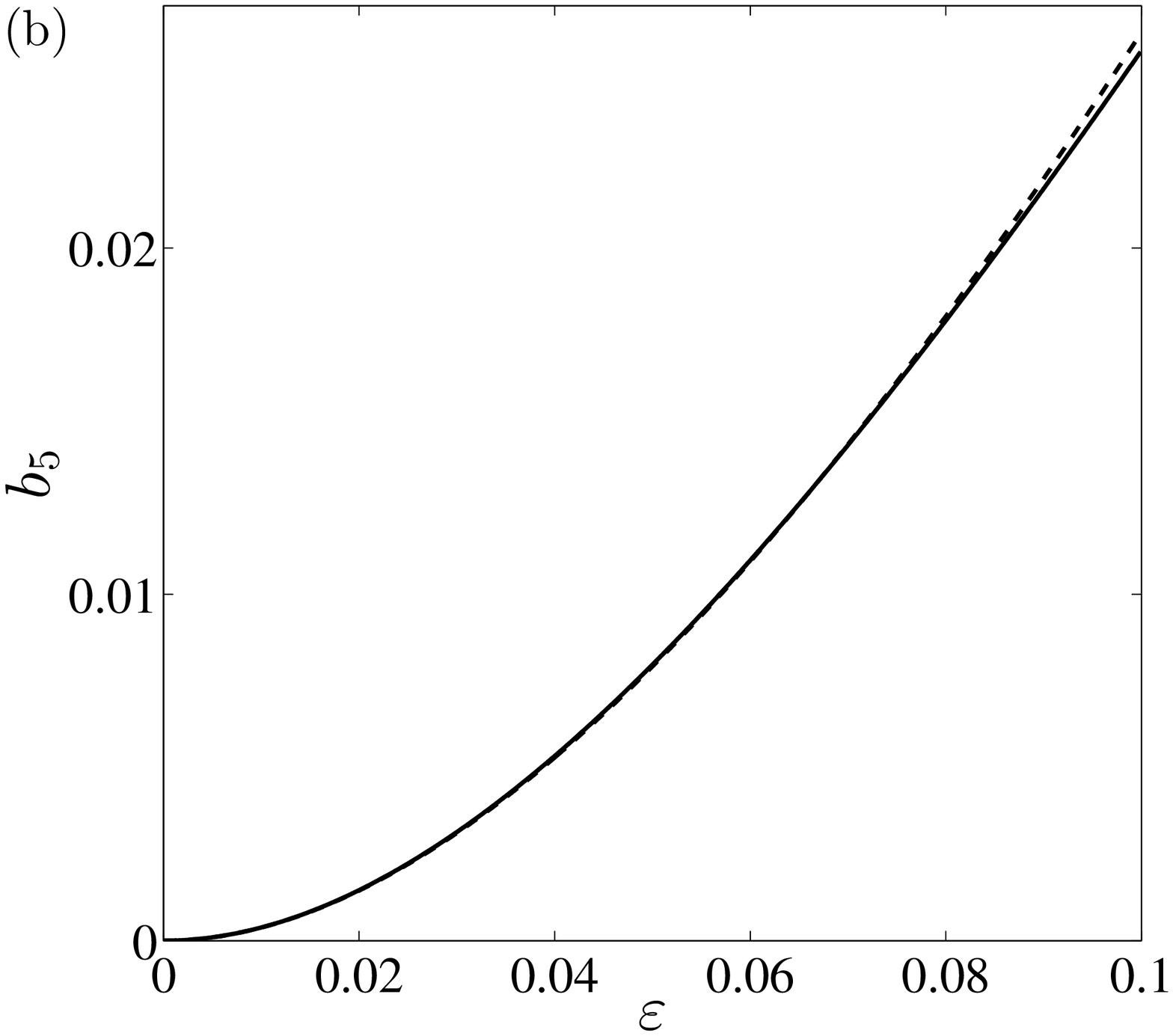}
\includegraphics[trim = 10mm 10mm 5mm 10mm,width=0.39\textwidth]{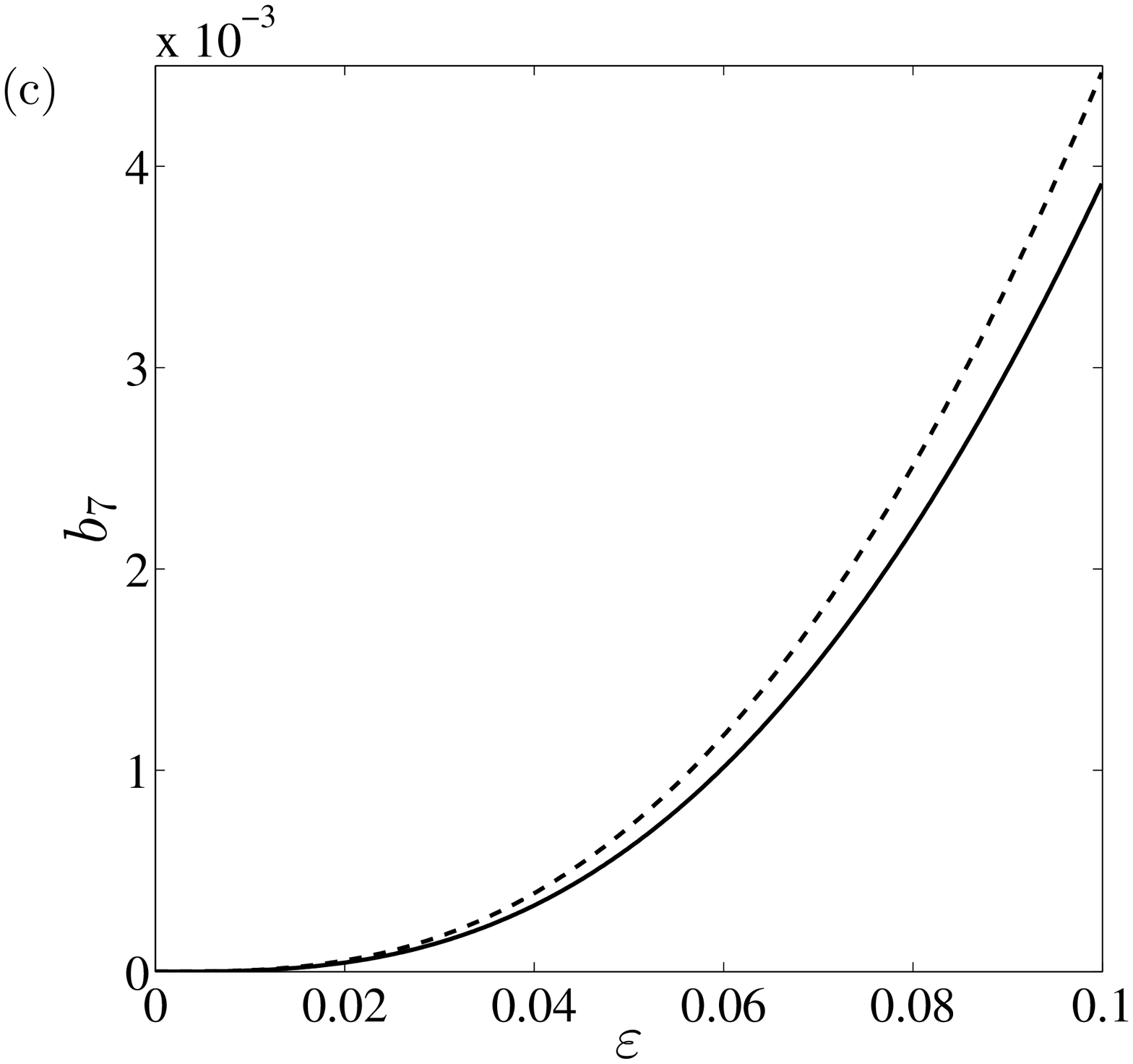}}
\par\end{centering}
\caption{\label{b3}Parameters $b_3$ (a), $b_5$ (b) and $b_7$ (c) as a function of $\varepsilon$. Solid lines: analytical results, dashed lines: approximation through Eq.~(\ref{compactformula}).}
\end{figure*}

\noindent The optimal values of $b_i$, for $i=3$, 5 and 7, obtained through Eq.~(\ref{bi}), are depicted in Fig.~\ref{b3}. This figure also shows that the regression\begin{equation}
b_{i,\text{opt}}=\frac{\left(2\varepsilon\right)^{\frac{i-1}{2}}}{1+3.5\times 1.5^{\frac{i-3}{2}}\varepsilon}\label{compactformula}
\end{equation}
provides an excellent approximation to the analytical results for $b_3$ and $b_5$, and a slight overestimate for $b_7$. Considering the dimensional parameters, Eq.~(\ref{compactformula}) becomes 
\begin{equation}
k_{2i}=\frac{\left(2\varepsilon\right)^{\frac{i-1}{2}}\varepsilon}{1+3.5\times 1.5^{\frac{i-3}{2}}\varepsilon}k_{1i}\label{k2i}
\end{equation}
Thanks to its simplicity, Eq.~(\ref{k2i}) can be used to calculate rapidly the optimal values of the nonlinear springs of the NLTVA. Together with Eq.~(\ref{DHrule2}), they offer a complete design of the NLTVA.

\subsection{Numerical validation}

\begin{figure*}
\begin{centering}
\makebox[\textwidth][c]{\includegraphics[trim = 10mm 10mm 10mm 10mm,width=0.39\textwidth]{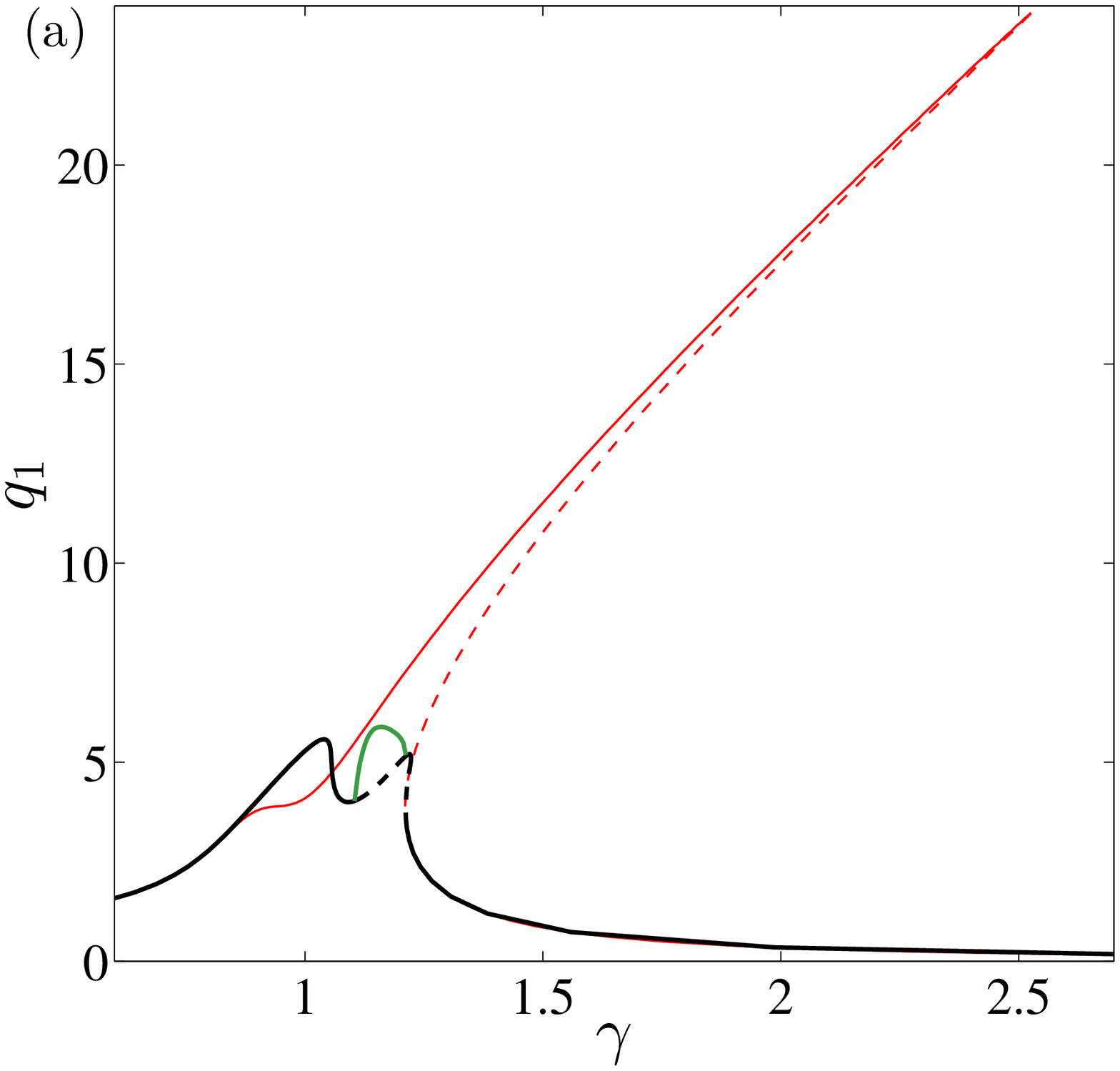}
\includegraphics[trim = 10mm 10mm 10mm 10mm,width=0.39\textwidth]{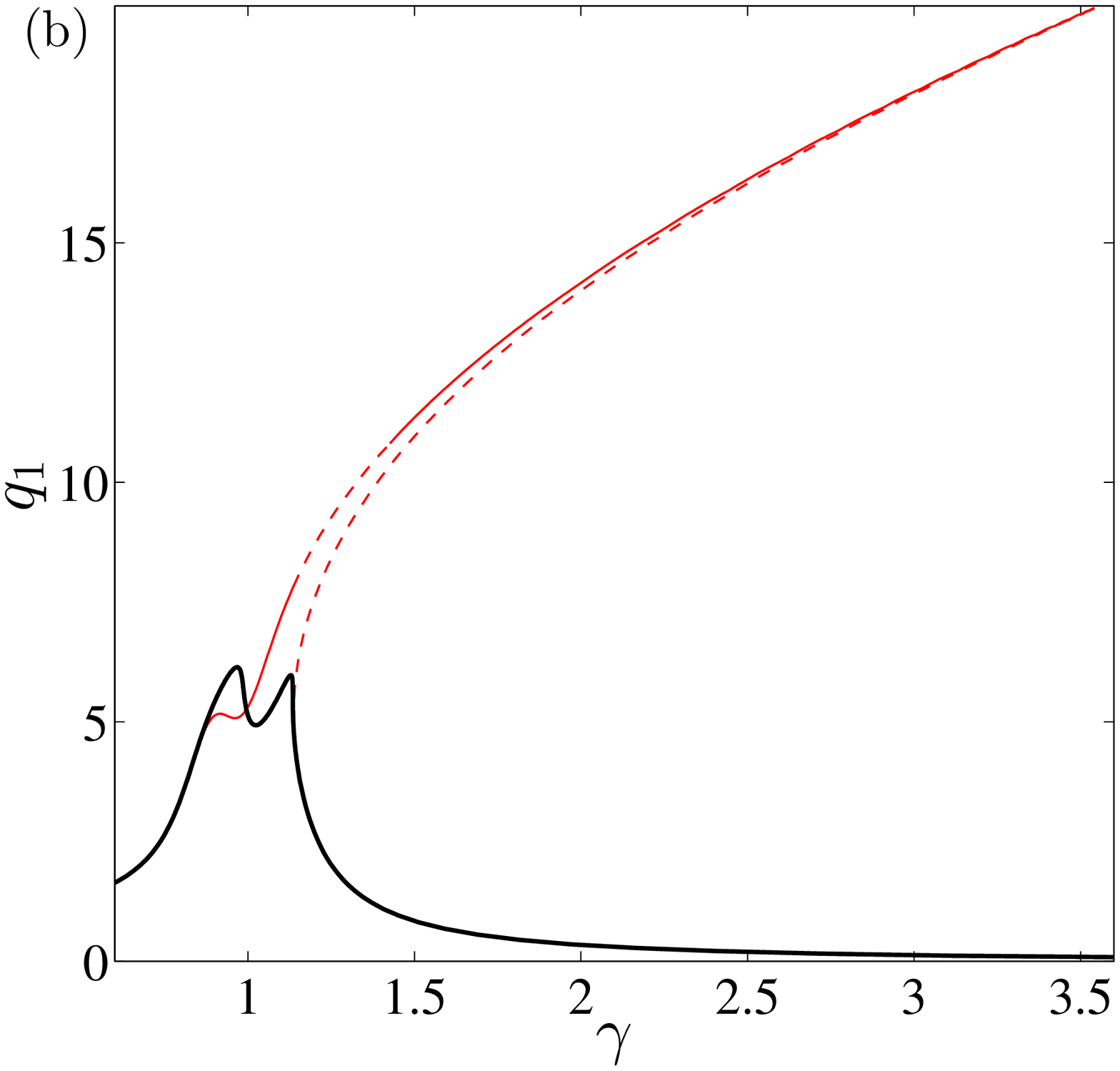}
\includegraphics[trim = 10mm 10mm 10mm 10mm,width=0.39\textwidth]{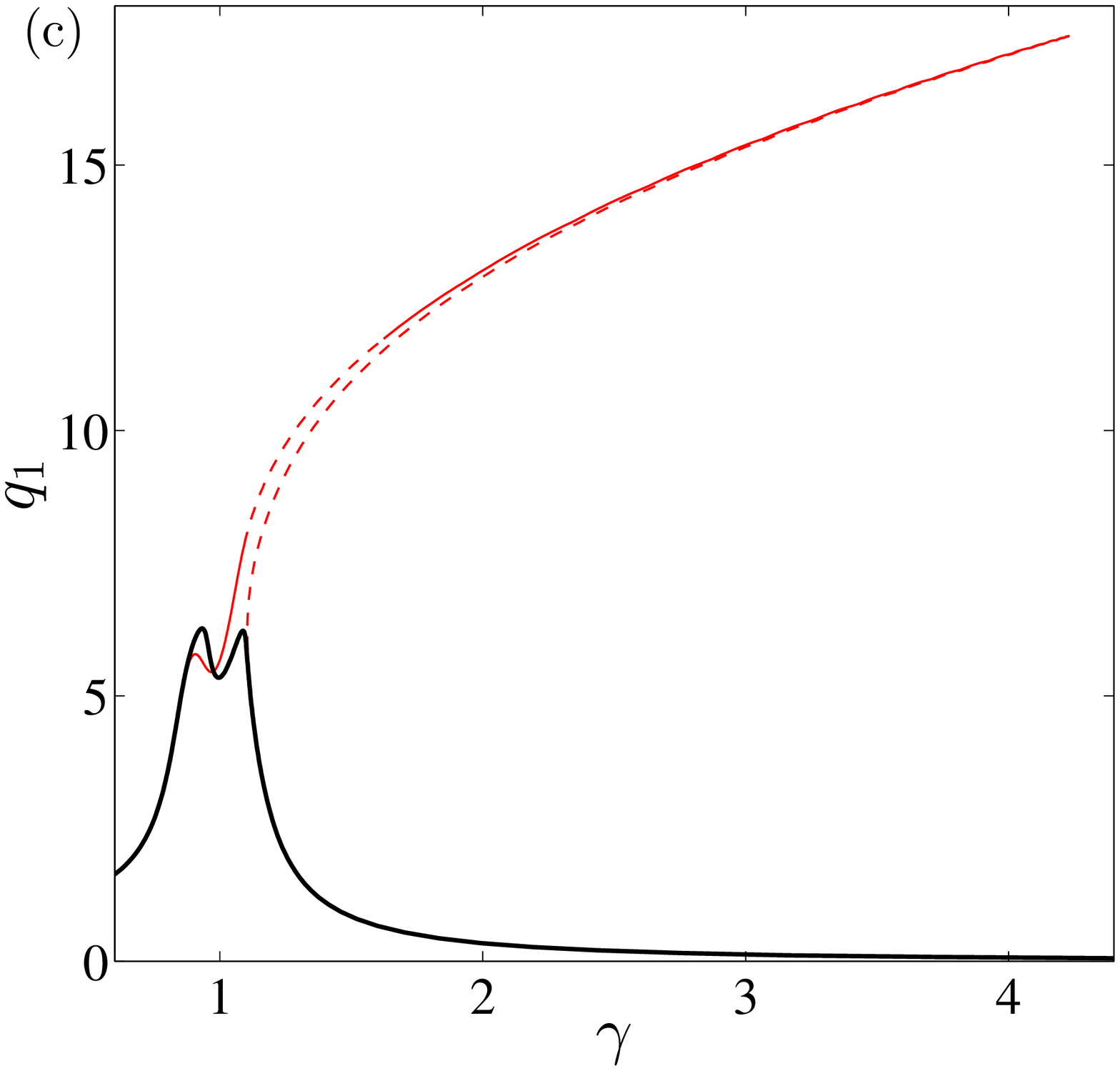}}
\par\end{centering}
\caption{\label{num_valid}{Frequency response of a primary system with cubic (a), quintic (b) or seventh-order (c) nonlinearity and an attached LTVA (red) or NLTVA (black). $\varepsilon=0.05$, $\alpha_3=0.013$ (a), $\alpha_5=1.3\times 10^{-4}$ (b), $\alpha_7=1.3\times 10^{-6}$ (c), whereas $b_3=0.0851$, $b_5=0.0079$ and $b_7=7.17\times10^{-4}$, according to Eq.~(\ref{compactformula}). Dashed lines indicates unstable solutions; green line indicates quasiperiodic solutions for the system with an attached NLTVA.}}
\end{figure*}

\begin{figure*}
\begin{centering}
\makebox[\textwidth][c]{\includegraphics[trim = 10mm 10mm 10mm 10mm,width=0.39\textwidth]{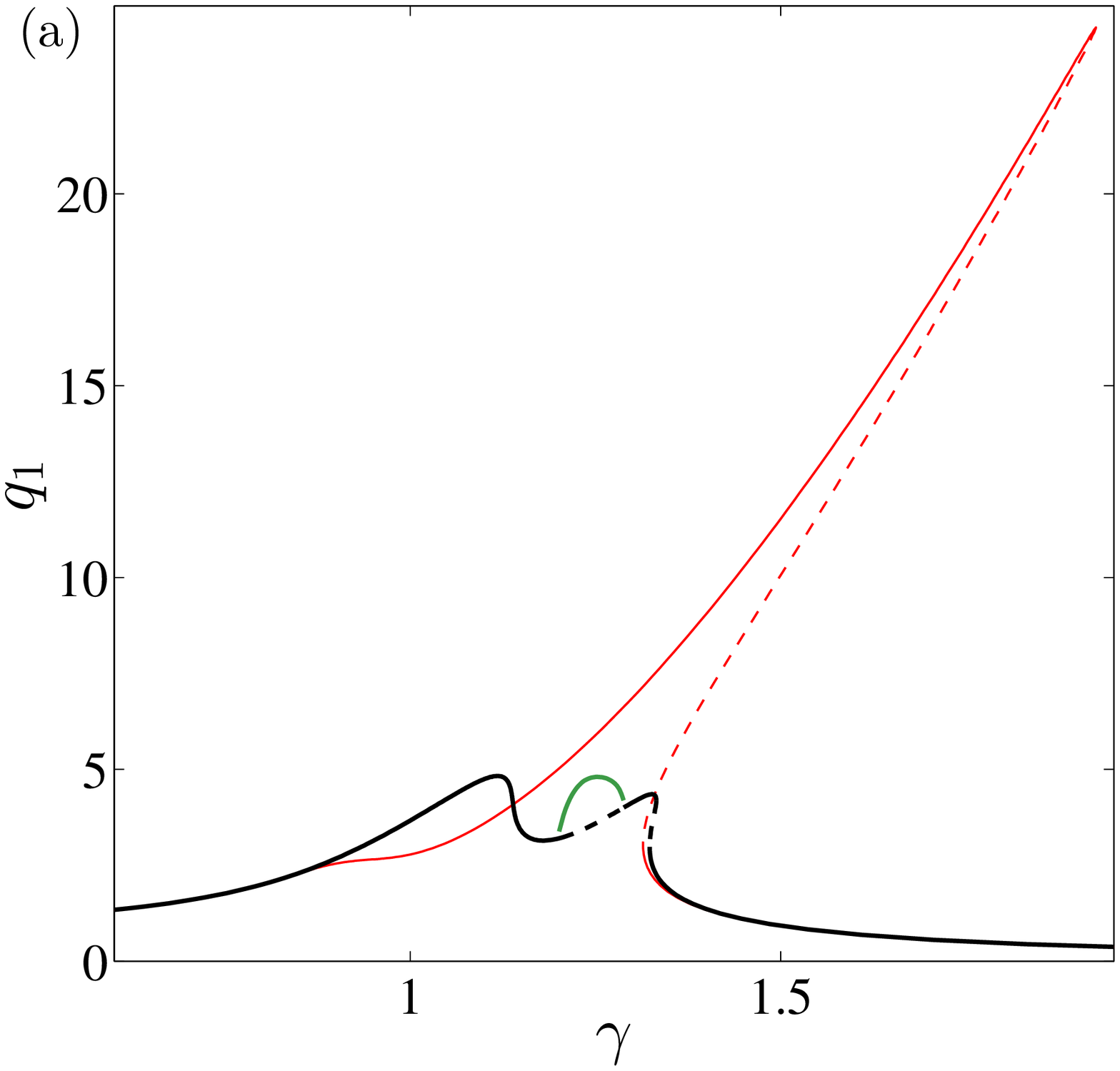}
\includegraphics[trim = 10mm 10mm 10mm 10mm,width=0.39\textwidth]{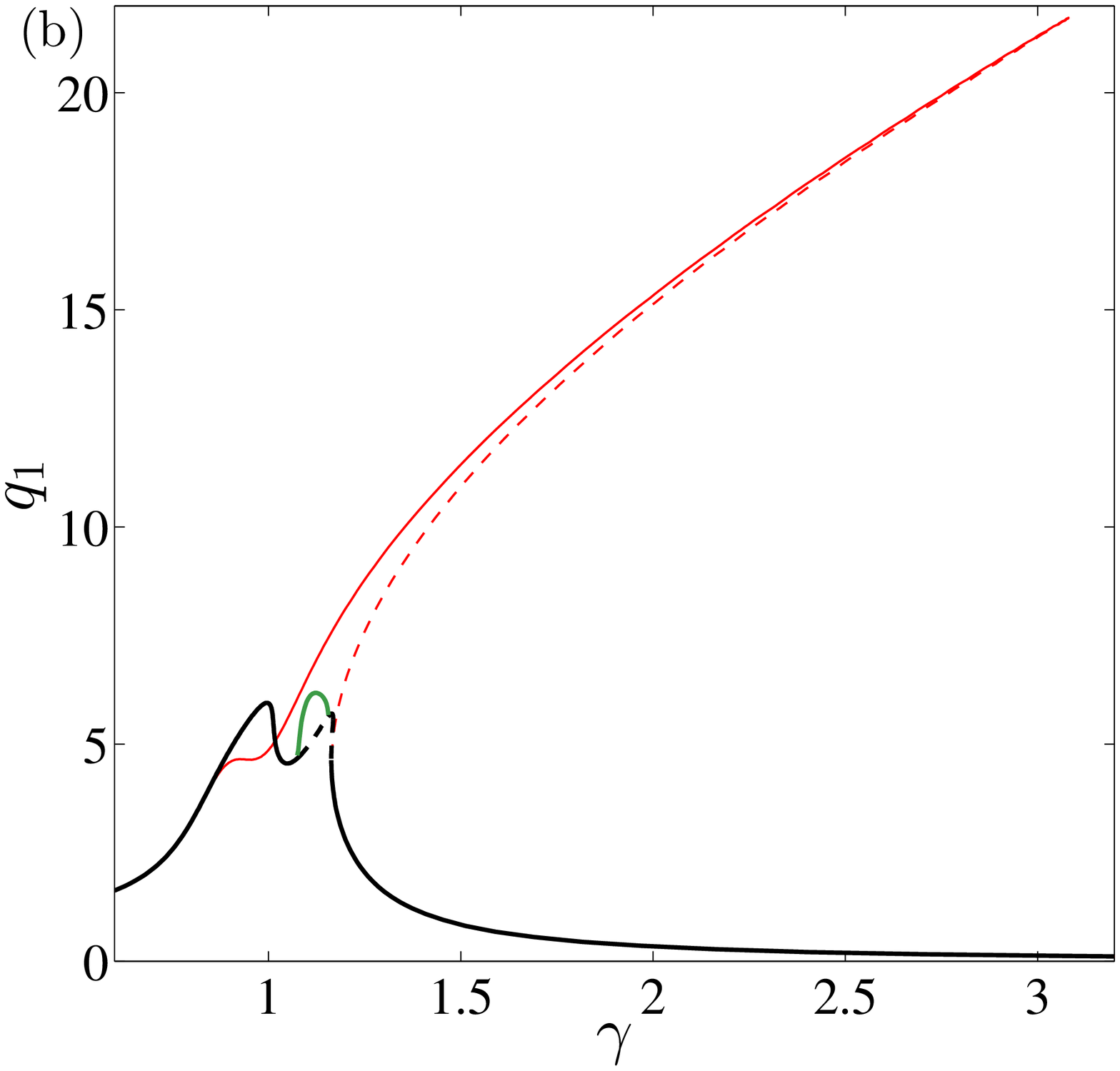}
\includegraphics[trim = 10mm 10mm 10mm 10mm,width=0.39\textwidth]{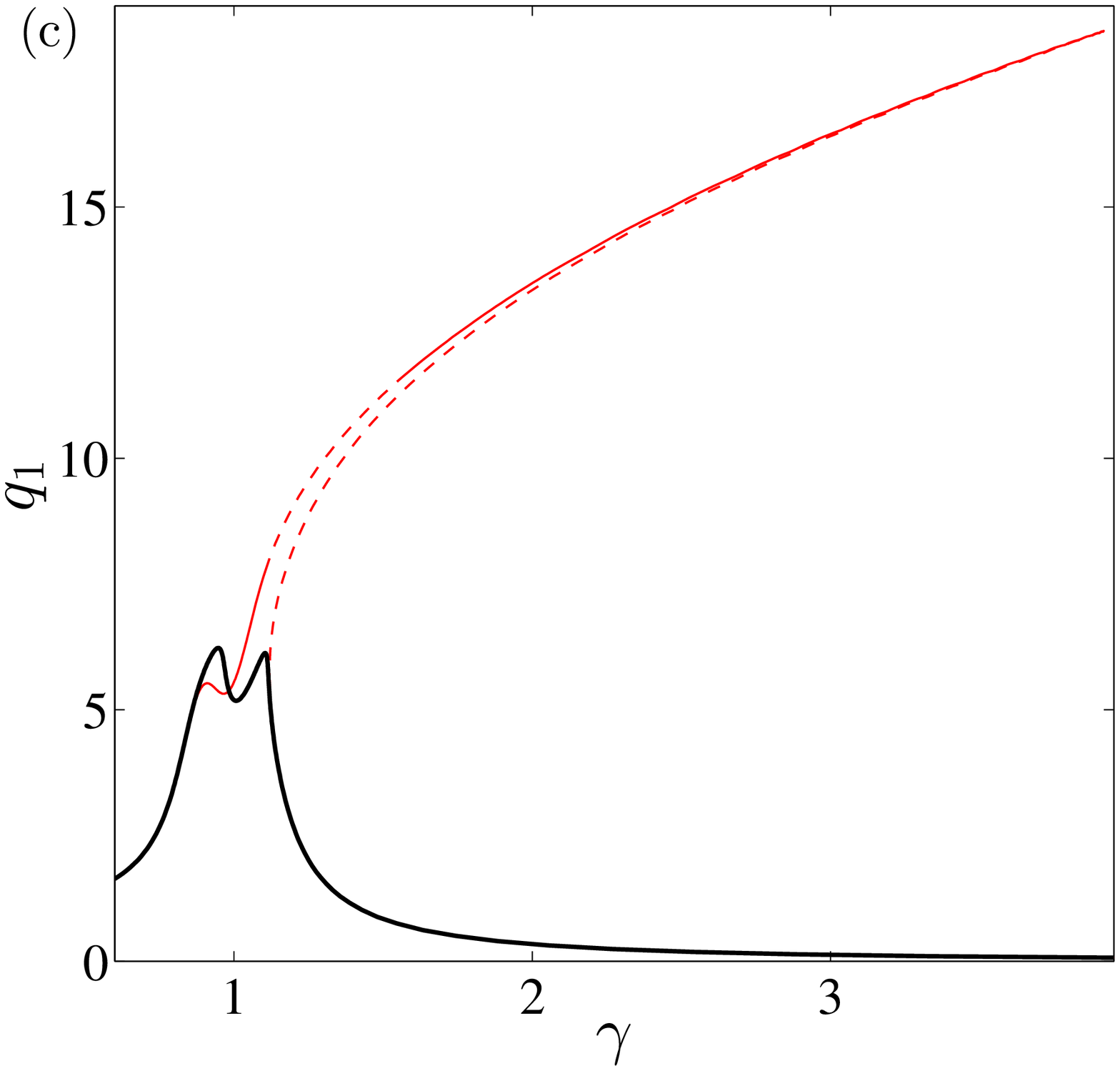}}
\par\end{centering}
\caption{\label{num_valid_pari}{Frequency response of a primary system with quadratic (a), quartic (b) or sixth-order (c) nonlinearity and an attached LTVA (red) or NLTVA (black). $\varepsilon=0.05$, $\alpha_2=0.13$ (a), $\alpha_4=1.3\times 10^{-3}$ (b), $\alpha_6=1.3\times 10^{-5}$ (c), whereas $b_2=0.2767$, $b_4=0.026$ and $b_6=0.0024$, according to Eq.~(\ref{compactformula}). Dashed lines indicates unstable solutions; green lines indicate quasiperiodic solutions for the system with an attached NLTVA.}}
\end{figure*}

\noindent Three primary systems with cubic, quintic and seventh-order nonlinearity, respectively, are first considered. The resulting frequency response curves are depicted in Figs. \ref{num_valid}.
These responses were computed using a path-following algorithm combining shooting and pseudo-arclength continuation similar to that used in \cite{Peeters}.
The red curves, referring to the nonlinear primary system with an attached LTVA ($\mathbf b=\mathbf 0$), clearly illustrate that the linear absorber is ineffective in all considered cases.
Conversely, the NLTVA (black lines) is able to mitigate the resonant vibrations of the primary system very effectively. The amplitudes of the two resonance peaks are almost equal for all nonlinearities, which validates our analytical developments. The same conclusion can be reached for primary systems with even-degree polynomial nonlinearities, as confirmed in Fig.~\ref{num_valid_pari}. The formula (\ref{k2i}) is therefore valid for polynomial nonlinearities of any degree.

\begin{figure*}
\begin{centering}
\makebox[\textwidth][c]{\includegraphics[trim = 10mm 10mm 10mm 10mm,width=0.39\textwidth]{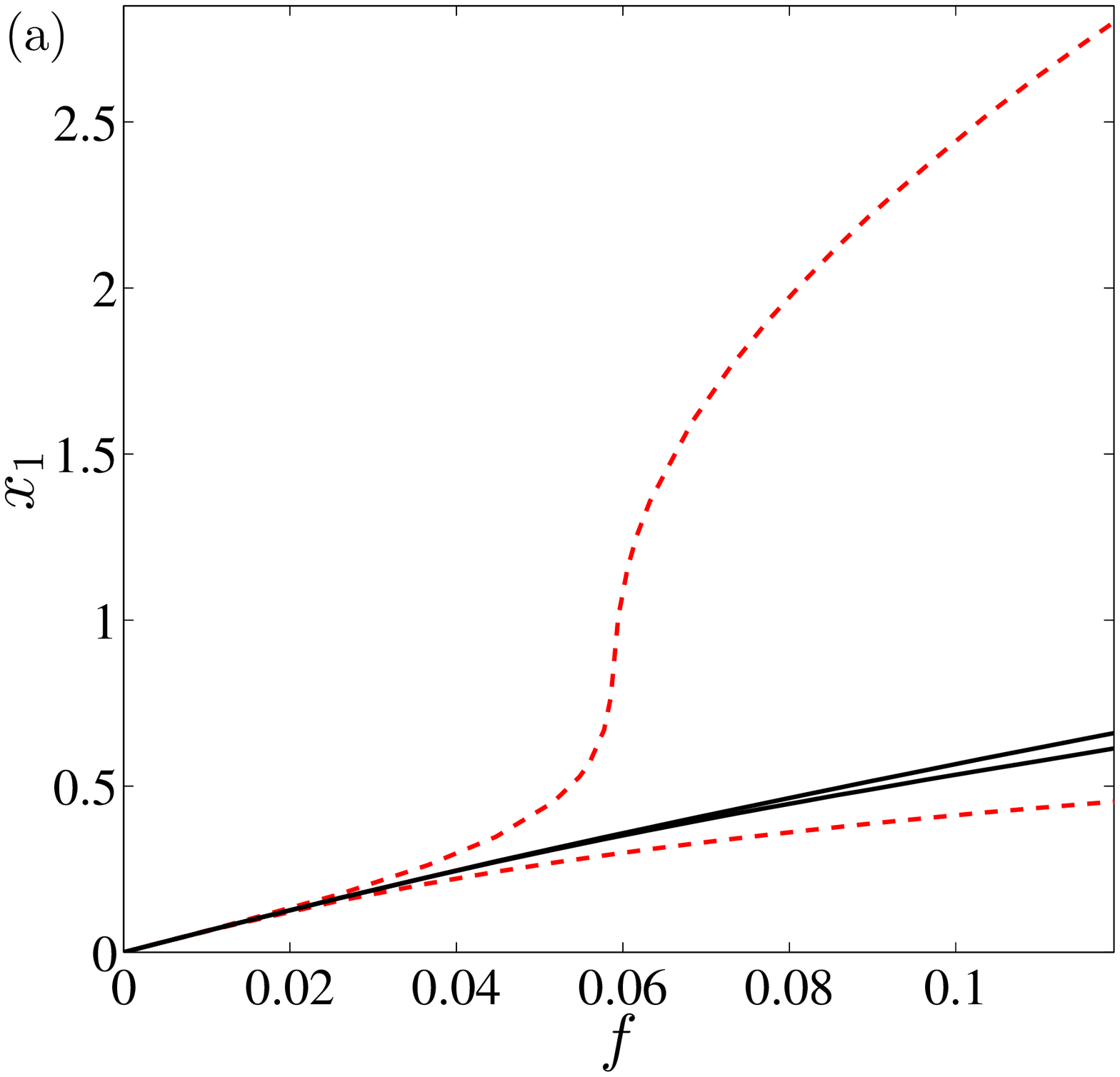}
\includegraphics[trim = 10mm 10mm 10mm 10mm,width=0.39\textwidth]{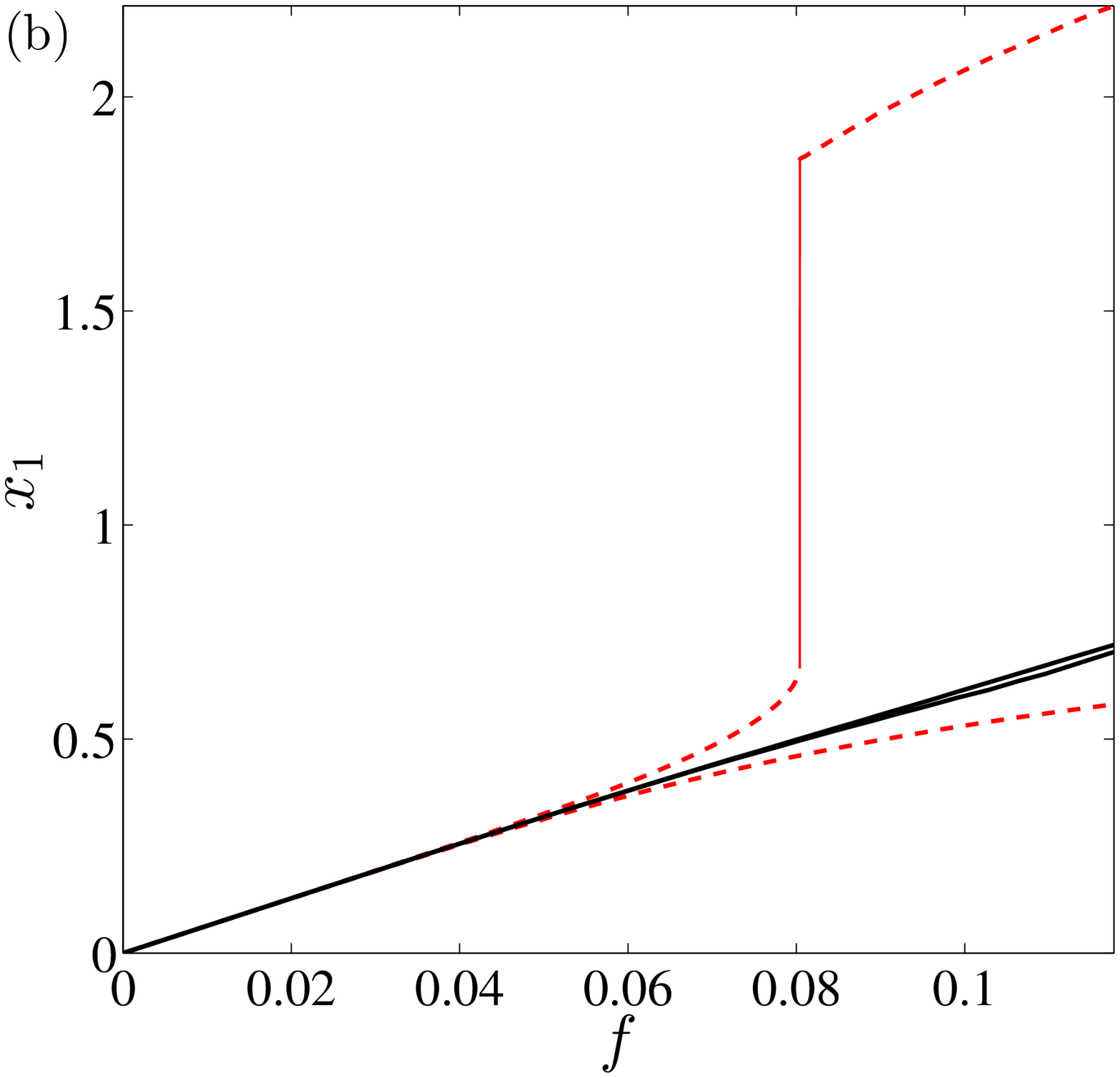}
\includegraphics[trim = 10mm 10mm 10mm 10mm,width=0.39\textwidth]{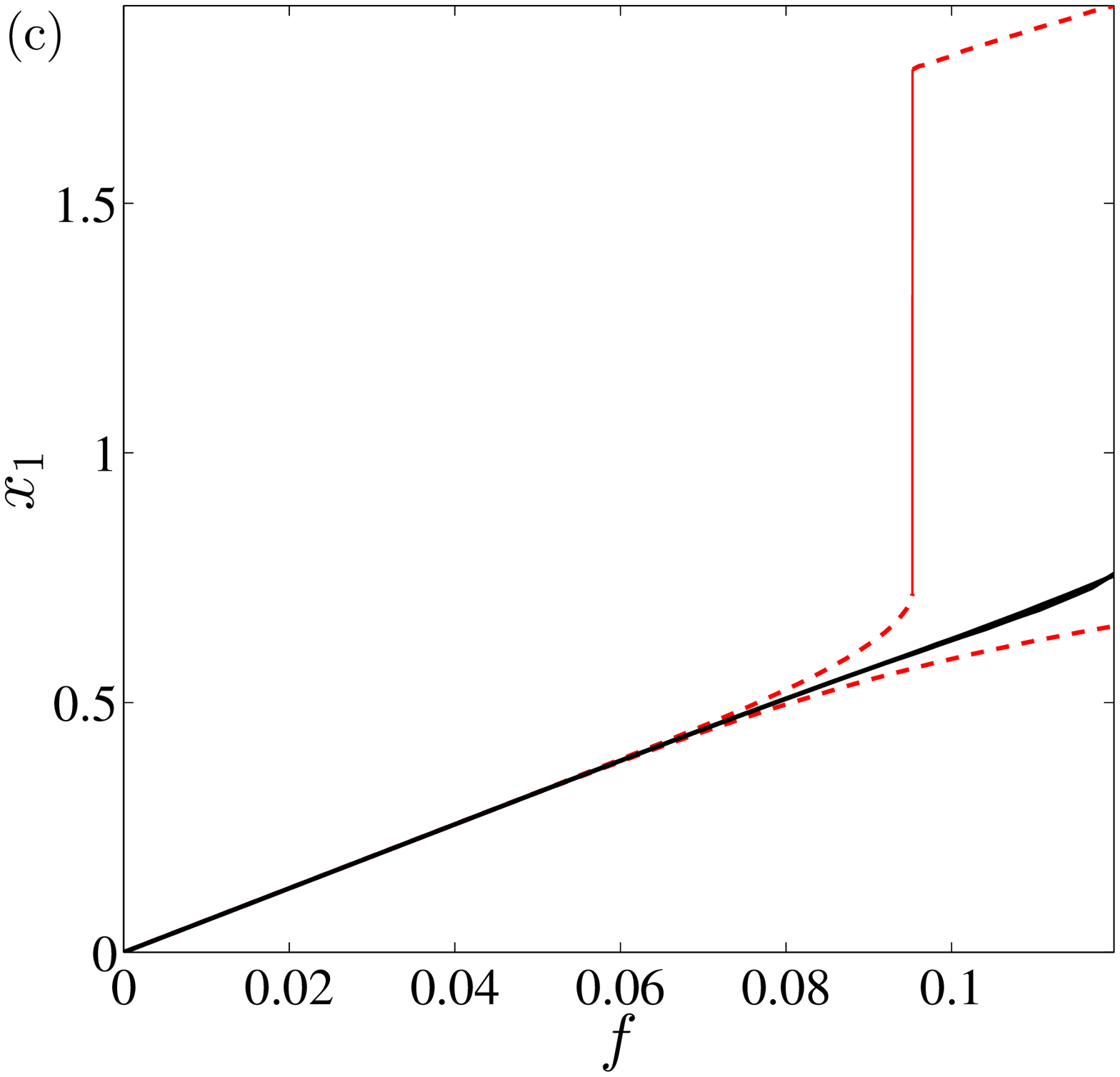}}
\par\end{centering}
\caption{\label{force_ampl}Peak amplitudes of a primary system with cubic (a), quintic (b) or seventh-order (c) nonlinearity and an attached LTVA (red) or NLTVA (black). $m_1=1$, $k_{11}=1$, $k_{13}=k_{15}=k_{17}=1$ and $\varepsilon=0.05$. The other parameters are chosen according to Eqs. (\ref{DHrule2}) and (\ref{k2i}).}
\end{figure*}

Figure \ref{force_ampl} illustrates the (dimensional) amplitude of the two resonant peaks for different values of the forcing amplitude and for nonlinearities of order 3, 5 and 7. For the LTVA, the two peaks rapidly diverge from each other, which confirms that this absorber is not effective for the mitigation of the considered nonlinear oscillations. In Figs. \ref{force_ampl}(b) and (c), the amplitude of one of the peaks undergoes a sudden jump which, as explained in \cite{Detroux}, is due to the merging of a detached frequency curve with the main frequency response curve.

For the NLTVA, the two peaks have approximately the same amplitude, which is the numerical evidence of the effectiveness of the proposed nonlinear equal-peak method. An important consequence of this result is that the range of validity of formulas (\ref{compactformula}) and (\ref{k2i}), which were developed under the assumption of small $\alpha_i$ (i.e., weak nonlinearity or, equivalently, weak forcing), extends to large values of $\alpha_i$ as well. This result, essential for the practical usefulness of the proposed tuning rule, is due to the adoption of the principle of similarity. A remarkable feature of these results is also the seemingly linear relation between the response amplitude and the forcing amplitude. This observation seems to suggest that the addition of a properly-designed nonlinear component in a nonlinear system can, to some extent, linearize the dynamics of the coupled system.

In Figs.~\ref{num_valid}(a) and \ref{num_valid_pari}(a,b) the frequency response of the system with the NLTVA presents unstable portions between the two resonant peaks.
The instability is due to a pair of Neimark-Sacker bifurcations which generate a branch of quasiperiodic oscillations (green lines).
We note that the corresponding amplitudes are not significantly larger than the two peaks, thus it does not compromise the effectiveness of the NLTVA.


\begin{figure*}
\begin{centering}
\makebox[\textwidth][c]{\includegraphics[trim = 10mm 10mm 10mm 10mm,width=0.39\textwidth]{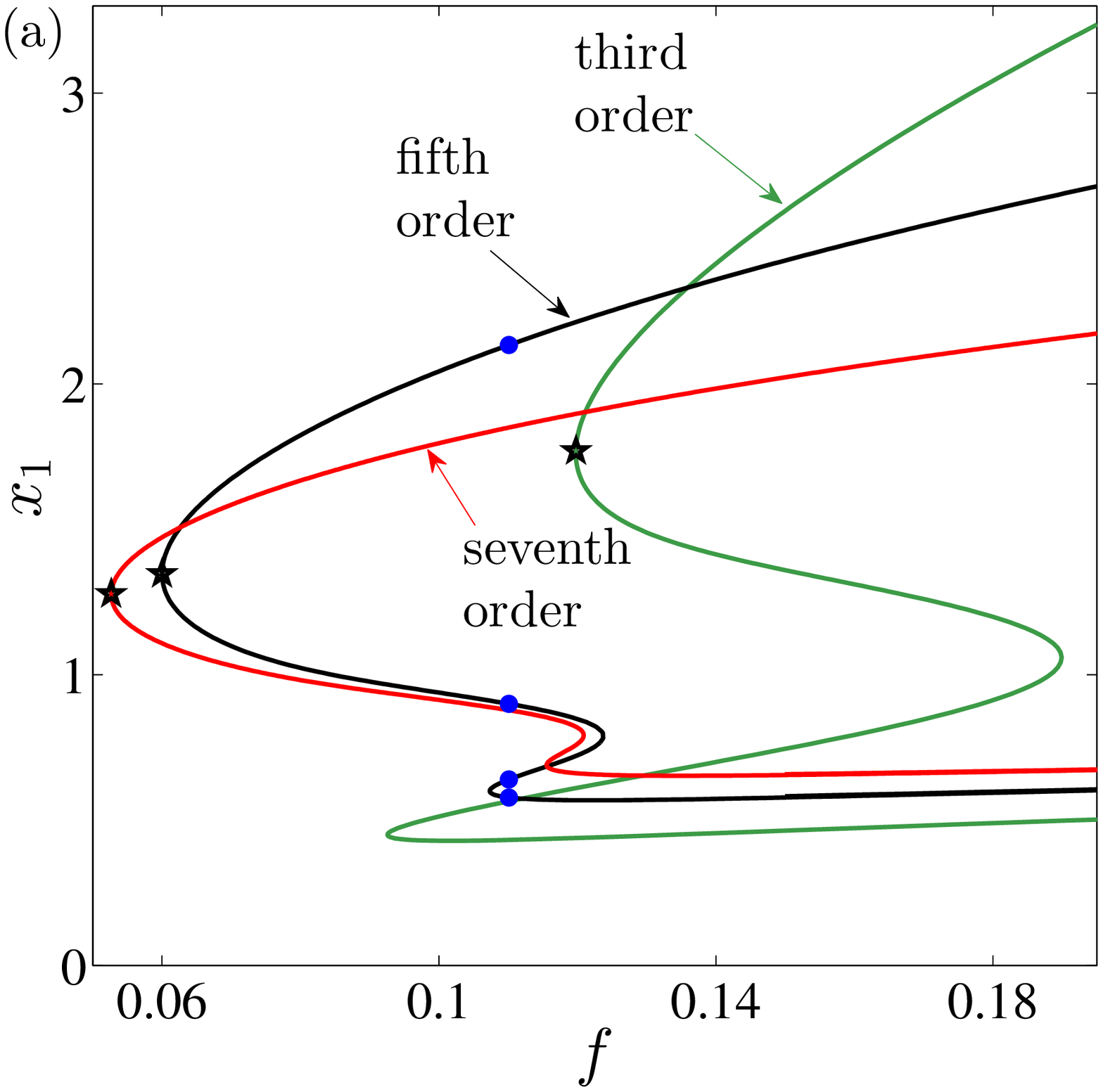}
\includegraphics[trim = 10mm 10mm 10mm 10mm,width=0.39\textwidth]{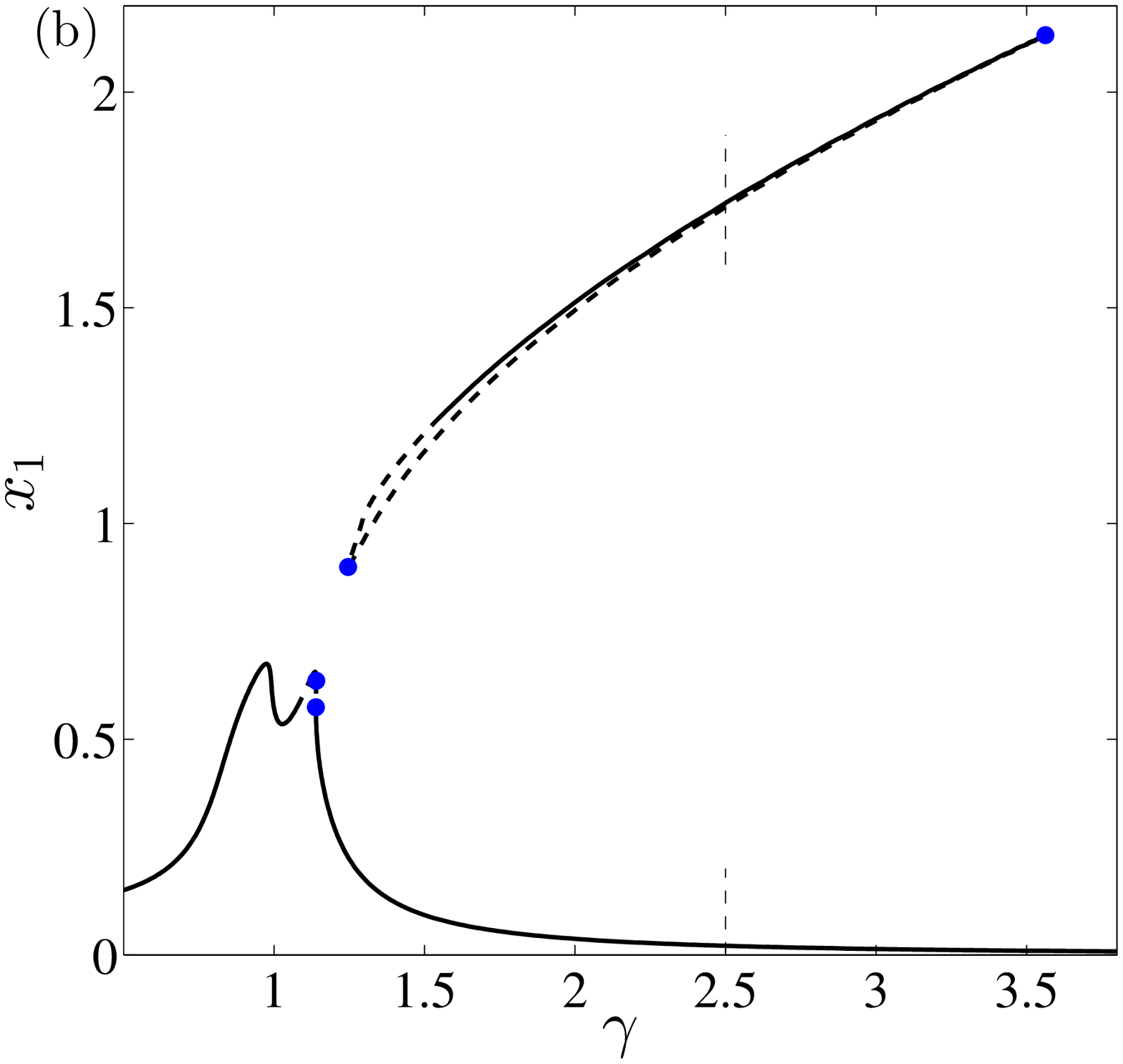}
\includegraphics[trim = 10mm 10mm 10mm 10mm,width=0.39\textwidth]{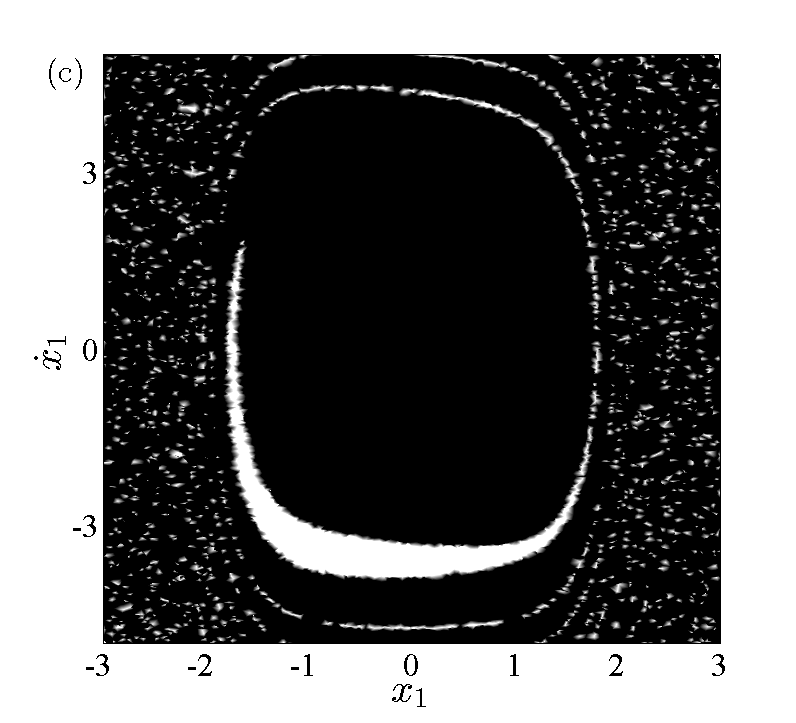}}
\par\end{centering}
\caption{\label{isolas}(a) Loci of fold bifurcations for DRCs detection; green, black and red curves refer to third-, fifth- and seventh-order nonlinearity, respectively, of primary systems with attached NLTVA; during the evaluation $m_1=1$, $k_{11}=1$, $k_{13}=k_{15}=k_{17}=1$ and $\varepsilon=0.05$. (b) Frequency response of a primary system with quintic nonlinearity, $f=0.11$. (c) Basin of attraction (white area) of the DRC for $\gamma=2.5$, $x_2=x_2'=0$. Black stars in (a) correspond to the onset of DRCs; blue dots in (a) and (b) mark the tracked fold bifurcations.}
\end{figure*}

Another source of detrimental dynamics is the appearance of detached resonant curves (DRCs).
The onset of DRCs can be detected by tracking the fold bifurcations that limit their domain of existence, as illustrated in Fig.~\ref{isolas}(a) for a system with third-, fifth- or seventh-order nonlinearity. Their appearance is marked by the merging of two branches of fold bifurcations, indicated by black stars in the figure. 
The three curves of Fig.~\ref{isolas}(a), obtained adopting a harmonic balance technique \cite{detroux2015}, show that for greater polynomial degree DRCs appear at lower forcing amplitudes. This represents a risk for the effectiveness of the absorber, since the DRCs have large amplitudes.
However, DRCs are not very robust, as proven by the basin of attraction illustrated in Fig.~\ref{isolas}(c) for $f=0.11$ and $\gamma=2.5$.
Furthermore, as illustrated in Fig.~\ref{isolas}(b), they present large portions of 
unstable motion.
The interested reader can refer to \cite{HabibMSSP,Detroux} for further details.


\section{Additivity of nonlinearities}\label{addit}

\noindent We now consider a primary system comprising third-, fifth- and seventh-order nonlinear terms. The dimensional parameters of the primary system are $m_1=1$, $k_{11}=1$, $k_{13}=1$, $k_{15}=1$ and $k_{17}=1$, such that $\alpha_3=f^2$, $\alpha_5=f^4$ and $\alpha_7=f^6$. The absorber has a mass $m_2=0.05$, thus $\varepsilon=0.05$.
$\lambda$ and $\mu_2$ are chosen according to Eq.~(\ref{DHrule2}), i.e., $\lambda=0.9524$ and $\mu_2=0.1339$.
Considering that the peaks of the dimensionless underlying linear system are such that $q_1\sim 10$, for any value of $f$, the nonlinear forces $\alpha_3q_1^3$, $\alpha_5q_1^5$ and $\alpha_7q_1^7$ have comparable amplitudes and participate in the dynamics to a similar extent.

\begin{figure*}
\begin{centering}
\makebox[\textwidth][c]{\includegraphics[trim = 10mm 10mm 10mm 10mm,width=0.39\textwidth]{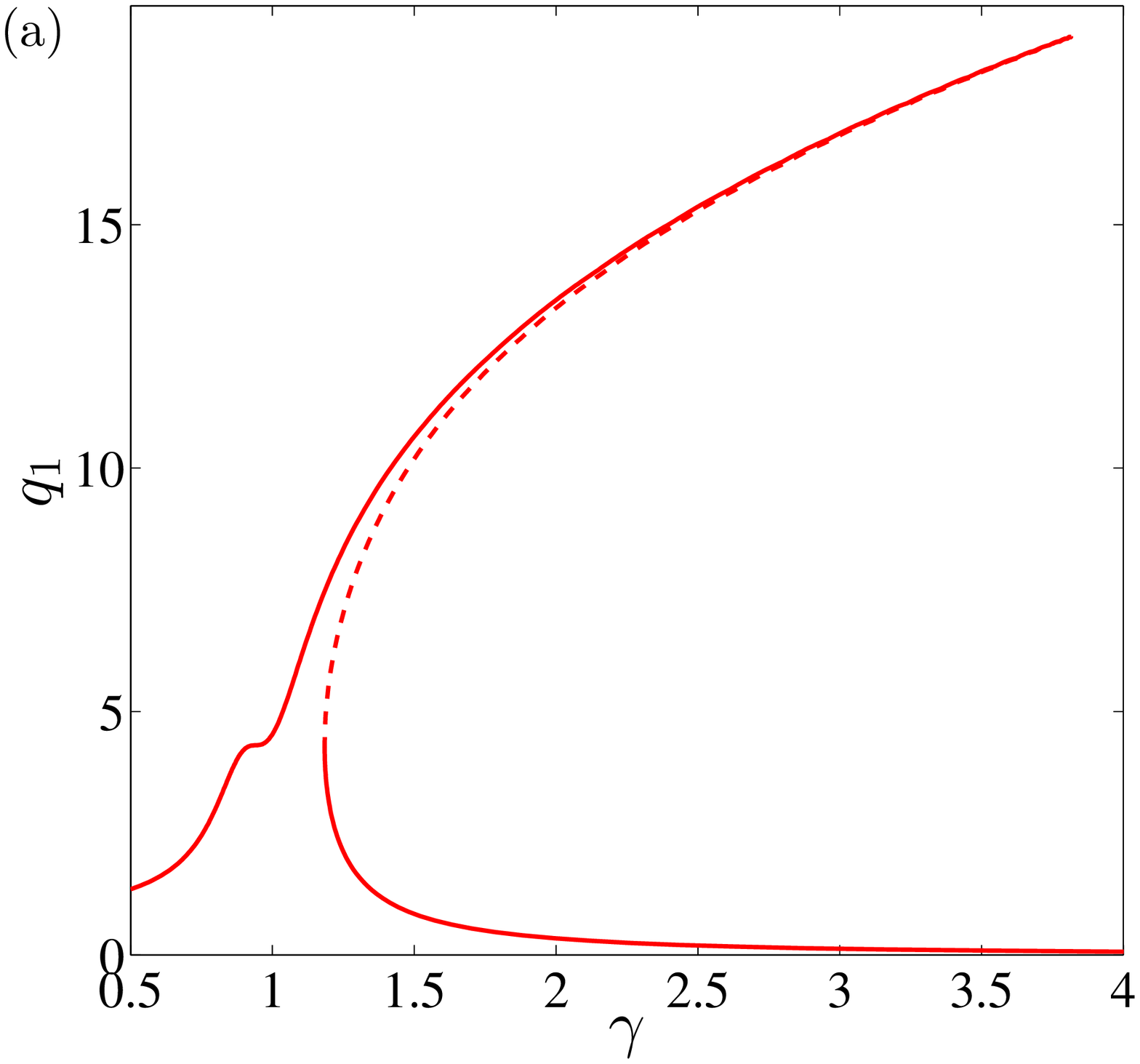}
\includegraphics[trim = 10mm 10mm 10mm 10mm,width=0.39\textwidth]{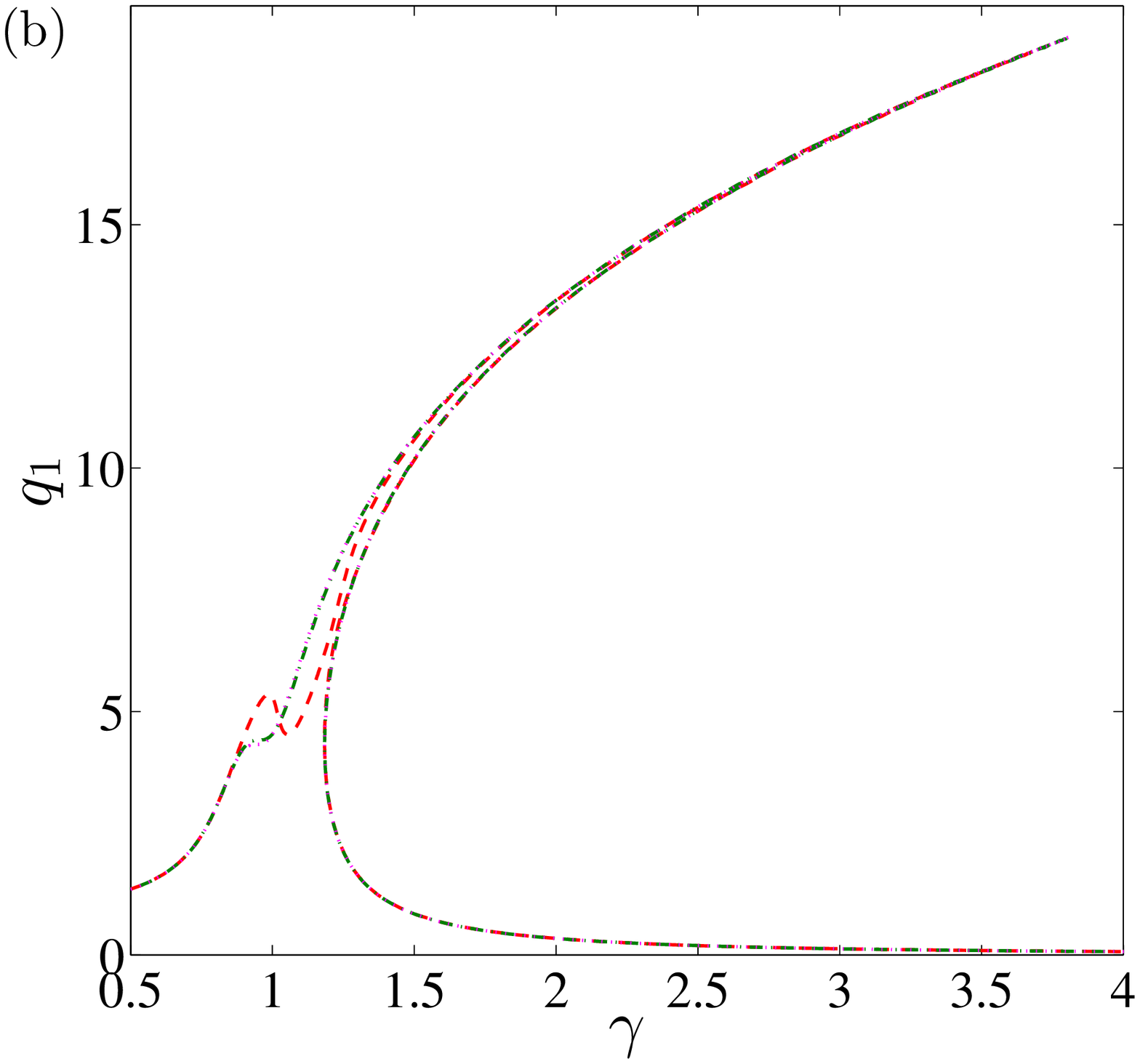}
\includegraphics[trim = 10mm 10mm 10mm 10mm,width=0.39\textwidth]{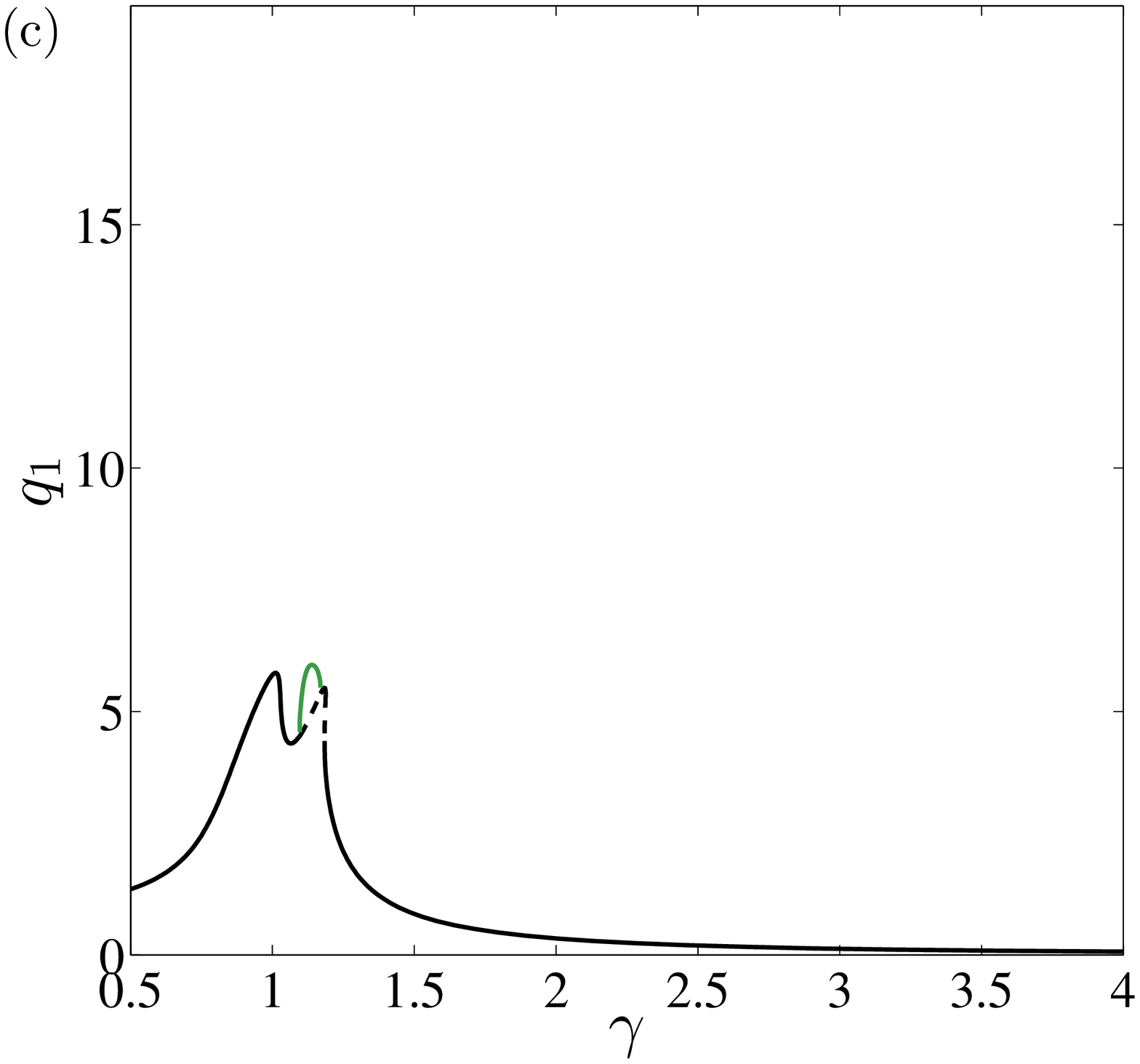}}
\par\end{centering}
\caption{\label{case_study}{Frequency response of a primary system comprising third-, fifth- and seventh-order nonlinearities coupled to different absorbers for $f=0.085$. (a) LTVA; (b) NLTVA with a single nonlinearity of either third (red dashed line), fifth (green dash-dotted line) or seventh (magenta dotted line) order; (c) complete NLTVA including all three nonlinearities. $b_3=0.0851$, $b_5=0.0079$ and $b_7=7.17\times10^{-4}$, according to Eq.~(\ref{compactformula}). (a) and (c) dashed lines indicates unstable solutions; (c) green line indicates quasiperiodic solutions.}}
\end{figure*}

Figure \ref{case_study} depicts the frequency response of the primary system coupled to different absorbers, namely a LTVA, a NLTVA with a single nonlinearity of either third, fifth or seventh order, and a NLTVA comprising all three nonlinearities, for $f=0.085$. Comparing the respective performance of the absorbers, it is immediately recognizable that the LTVA or the NLTVA with a single nonlinear component are practically ineffective, whereas the complete NLTVA successfully mitigates the resonant vibrations, in a way that resembles the underlying linear system.
As in Figs.~\ref{num_valid}(a) and \ref{num_valid_pari}(a,b), a branch of quasiperiodic motion exists and possesses an amplitude similar to that of the resonance peaks.

\begin{figure*}
\begin{centering}
\includegraphics[trim = 10mm 10mm 10mm 10mm,width=0.4\textwidth]{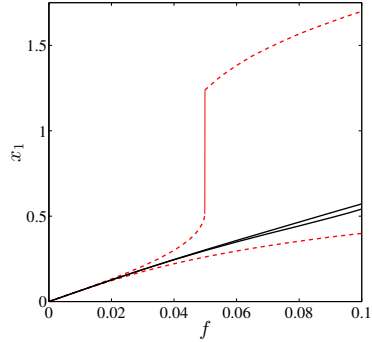}
\par\end{centering}
\caption{\label{force_case_study}Peak amplitudes for increasing values of the forcing amplitude $f$. Red: LTVA; black: complete NLTVA.}
\end{figure*}

Figure \ref{force_case_study} plots the amplitudes of the resonant peaks for increasing values of the forcing amplitude. Qualitatively similar results as those observed in Fig.~\ref{force_ampl} for a primary system with a single nonlinearity are obtained. Thanks to the property of additivity highlighted in Eq.~(\ref{bi}), the proposed nonlinear equal-peak method therefore extends as well to primary systems possessing multiple polynomial nonlinearities.

\section{Conclusions}

\noindent The fundamental principle of the NLTVA is the principle of similarity, which states that the absorber should possess the same nonlinearities as in the primary system. Relying on this principle, the objective of this paper was to derive analytically a tuning rule for extending the equal-peak method, which is widely used for the design of linear absorbers, to nonlinear systems. Eventually, we obtained a compact formula valid for any polynomial nonlinearity that can be used to rapidly design a NLTVA. Another interesting theoretical result of this study is the property of additivity of different nonlinearities, i.e., if different polynomial nonlinearities are present in the primary system, the polynomial nonlinearities in the NLTVA can be designed independently of each other. Throughout the paper, the NLTVA exhibited excellent performance and always outperformed the LTVA, something which is not often verified for nonlinear vibration absorbers \cite{Book,Ema}.

\section*{ACKNOWLEDGEMENT}

\noindent The authors gratefully acknowledge the financial support of the European Union (ERC Starting Grant NoVib 307265) and Javier Gonz\'alez Carbajal for fruitful discussions.

\section*{References}




\end{document}